# PULSES: multi-agent AI for autonomous excited-state model building and dynamics

Fangning Ren,[#] Lechen Dong,[#] Xu Chen, Weilin Zou, Bushra Alam, Fang Liu[*]
Department of Chemistry, Emory University, Atlanta, GA, 30322, USA

**Abstract** Simulating molecular excited-state processes requires resolving electronic and nuclear dynamics, but the cost of direct quantum-chemical calculations limits accessible system sizes and timescales. Machine-learning models can extend these simulations, yet they often need to be tailored to specific molecular systems and scientific objectives. Building and deploying such models requires coordinated decisions across conformational sampling, quantum-chemical labelling, model training, validation and dynamics. Here we introduce Paired agents for Unified Learning and Simulation of Excited States (PULSES), a multi-agent system that automates this workflow within expert-approved boundaries. A human expert and planner define the scientific objective, methods and decision boundaries, while an orchestrator coordinates domain agents and artifact-specific scientific assurance during execution. We evaluate PULSES on 15 computational chemistry tasks and demonstrate end-to-end applications to perylene exciton Hamiltonians, solvent-dependent biphenyl relaxation, and excited-state proton transfer in 10-hydroxybenzo[h]quinoline. These results demonstrate traceable, expert-bounded construction and deployment of specialized excited-state models, while retaining scientific validation as a requirement rather than assuming that successful execution establishes correctness.

## 1. Introduction

Electronic excited states govern how molecules absorb, transport, and transform energy, underpinning photochemistry, spectroscopy, light harvesting, and optoelectronic function. Predictive simulations of these processes must resolve not only excited-state energetics but also the nuclear motion and electronic-state transitions that determine experimentally observable outcomes [1]. Direct quantum-chemical dynamics can provide this information, but its cost restricts the accessible system sizes, environments, timescales and trajectory ensembles [2].

[*] Email: fang.liu@emory.edu

[#] F.R. and L.D. contributed equally to this work.

Physics-informed machine-learning models offer a route around this computational bottleneck. In ground-state atomistic simulation, machine-learning interatomic potentials can approach first-principles accuracy while extending simulations to larger systems and longer timescales [3,4], and transferable and foundation-style potentials have expanded the chemical space accessible to a single model [5,6]. Achieving comparable generality for molecular excited states is much harder. Excited-state dynamics evolves on multiple coupled potential-energy surfaces, where state ordering and electronic character can change with nuclear geometry and where near-degeneracies and conical intersections may control the outcome [7]. Depending on the problem, a model may need to represent several adiabatic energies and forces, nonadiabatic couplings, diabatic Hamiltonian elements, transition properties or environment-dependent interactions. Generating these labels is substantially more expensive and method-dependent than obtaining ground-state energies and forces [8], and the configurations needed for training are often concentrated in dynamically important regions that cannot be sampled adequately from equilibrium structures alone.

Transferable excited-state architectures are emerging and have shown promising generalization within selected molecular datasets or related chemical families [9]. However, recent reviews identify transferability across chemical compound space and the reliable treatment of coupled excited-state surfaces as continuing challenges [10], and current excited-state models have not yet reached the breadth and generality of ground-state universal potentials. These advances therefore complement rather than eliminate the need for models tailored to a specific system and scientific objective. Building such a fit-for-purpose model requires researchers to define the molecular system and environment, target process or observable, relevant state manifold and electronic-structure level; sample informative configurations; generate and curate reference quantities; select a representation; train and validate the model; and connect it to an appropriate dynamics protocol (Fig. 1a). These decisions are coupled: a coverage gap or scientific inconsistency discovered during validation or dynamics may require changes to sampling, electronic-structure calculations or the model itself. Rather than relying exclusively on increasingly general potentials, a complementary strategy is therefore to make the construction of reliable, system-specific excited-state models substantially faster and more reproducible.

Agentic artificial intelligence offers a natural way to coordinate such adaptive workflows. For the present discussion, we distinguish three capability levels. At the first level, an agent uses established scientific tools through a natural-language interface. At the second, it plans and executes a multi-step computational workflow, adapting tool selection and recovering from execution errors [11]. At the third, it constructs a new scientific model for the task at hand: the agent generates training data, trains and validates the model, and then deploys that model to answer the scientific question. Prior systems illustrate the first two levels. Chemistry agents such as ChemCrow [12], MDCrow [13] and El Agente Q [14] have demonstrated increasingly capable tool use and workflow automation. Structured automation can also be achieved without a multi-agent architecture: AutoSolvateWeb, for example, uses a chatbot-guided cloud platform to configure and execute a predefined multistep workflow for explicit-solvent QM/MM simulation [15]. Early systems for automated model construction have begun to appear in adjacent ground-state and scientific machine-learning settings [16]. An end-to-end agentic system that builds and uses custom models for excited-state dynamics, however, remains largely unexplored.

Reaching this level of autonomy raises two distinct control problems. First, the system must determine when full autonomy should begin. Continuous human-in-the-loop execution sacrifices much of the efficiency that automation is intended to provide, whereas converting an underspecified scientific goal directly into a long and resource-intensive sequence of operations risks producing an output that is technically complete but does not answer the user's intended question. Second, once execution begins, the system must maintain scientific validity without requiring an expert to inspect every intermediate result. Agent-generated workflows can fail visibly, for example when a calculation crashes or an agent becomes trapped in an unproductive repair loop. More consequential are silent scientific failures: a calculation finishes and returns plausible numbers, but the state assignment is wrong, the sampling misses a critical region, units or conventions are inconsistent, the validation set leaks training data [17], or the chosen dynamics approximation is incompatible with the mechanism. Visible failures consume human time, whereas silent failures can propagate into apparently credible but incorrect conclusions. Scientific autonomy therefore requires both an explicit boundary between expert planning and autonomous execution and stage-specific checks that determine whether an artifact is fit to enter the next stage of the workflow.

Here we introduce PULSES (Paired agents for Unified Learning and Simulation of Excited States), a multi-agent system for the autonomous construction and application of system-specific excited-state models. PULSES separates expert-approved scientific planning from autonomous execution: a human expert and the Planner Agent establish the objective, methods and decision boundaries before the Orchestrator coordinates specialist agents. During execution, each critical Domain Agent is paired with a functionally separate Scientific Assurance Agent that evaluates the artifacts produced at that stage before they propagate downstream (Fig. 1b,c). This architecture is designed to reduce both stalled workflows and silent scientific failures.

We evaluate PULSES on component tasks and on three end-to-end applications that require different model representations and dynamical approximations. The system constructs and deploys a learned exciton Hamiltonian for perylene assemblies, investigates solvent-dependent excited-state structural relaxation in biphenyl, and models excited-state intramolecular proton-transfer dynamics in 10-hydroxybenzo[h]quinoline. To our knowledge, comparable excited-state dynamics simulations have not been reported for the biphenyl problem; the predicted solvent dependence is qualitatively consistent with measurements on a long-chain-substituted biphenyl derivative [18], while the difference between the simulated and experimental molecules limits a direct quantitative comparison. Together, these studies show how agentic AI can progress from operating existing software to constructing, checking and using fit-for-purpose scientific models. PULSES provides a general strategy for making specialized excited-state simulations more accessible without treating automation as a substitute for scientific validation.

## 2. Results

### 2.1 Expert-approved planning and assurance-gated execution

PULSES addresses these two control problems through complementary mechanisms: an expert-approved autonomy boundary and assurance-gated domain execution (Fig. 1b,c). The first determines when the system is permitted to act autonomously; the second determines whether the scientific artifact produced at each stage is fit to propagate downstream.

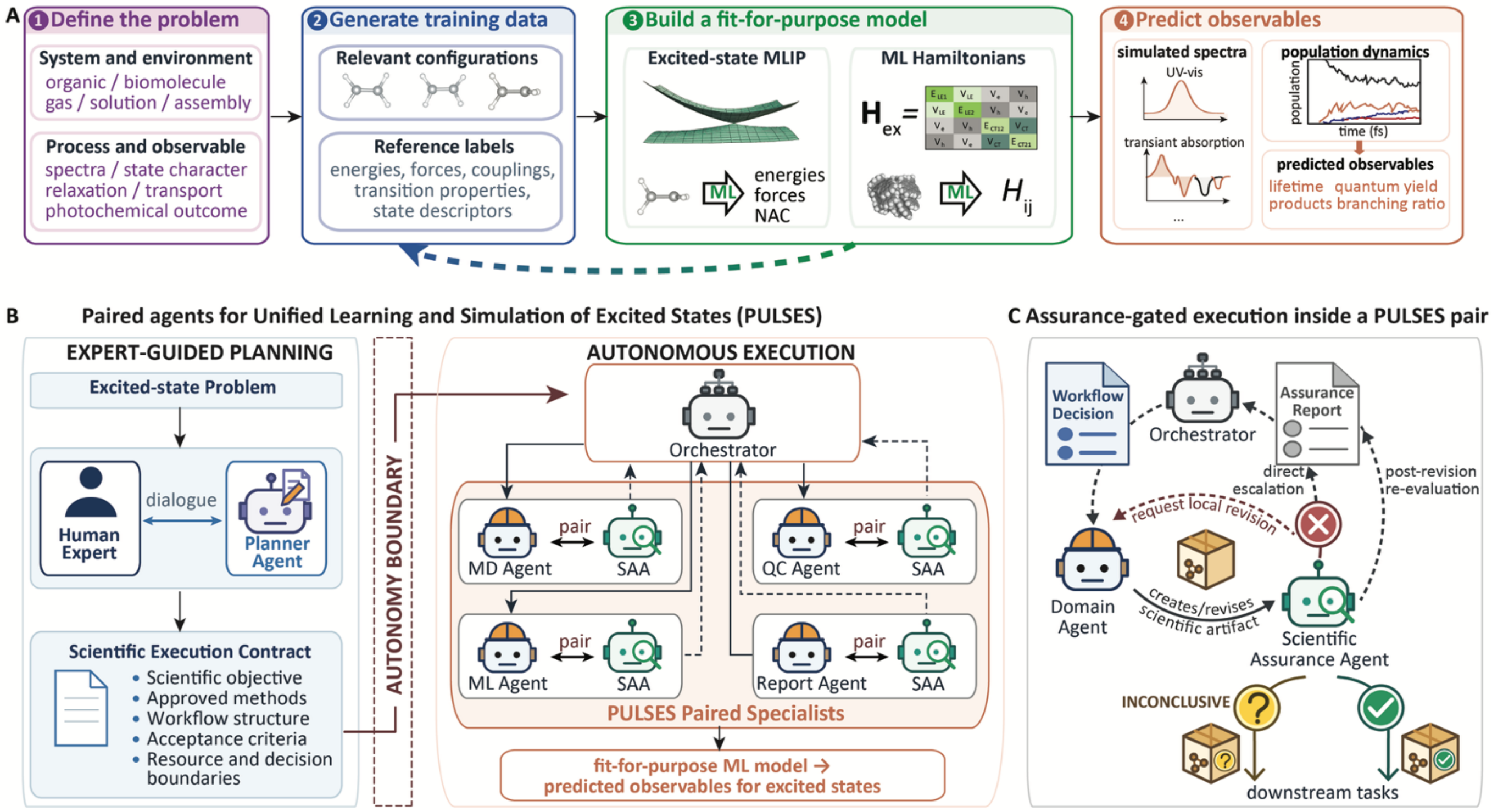

Fig. 1 From conventional fit-for-purpose excited-state model construction to expert-bounded, assurance-gated autonomy in PULSES. **a** A conventional expert-directed workflow defines an excited-state problem by its system, environment, process, target observable and electronic-structure level. These choices determine the configurations and reference labels used to train an excited-state machine-learning interatomic potential or Hamiltonian, which is then deployed to predict spectra, population dynamics and derived observables. The dashed return path denotes iterative refinement of the training data. **b** In PULSES, the Human Expert and Planner Agent agree on a Scientific Execution Contract during expert-guided planning. After the autonomy boundary, the Orchestrator coordinates domain-specific PULSES pairs, each comprising a Domain Agent and a Scientific Assurance Agent (SAA). **c** The Domain Agent creates or revises a scientific artifact, and the SAA assigns PASS, INCONCLUSIVE, or FAIL. Passed artifacts advance; inconclusive artifacts advance conditionally with a persistent warning for diagnostic use. Following failure, the SAA requests local revision or escalates directly; revised artifacts are re-evaluated, and the Orchestrator makes the workflow-level decision from the assurance report.

A human expert specifies the problem and iteratively discusses the proposed strategy with the Planner Agent, which translates the scientific objective into an executable plan. Their dialogue ends in a Scientific Execution Contract that records the objective, approved methods, workflow structure, acceptance criteria, and resource and decision boundaries. Execution does not begin until the expert approves this contract. The contract makes the autonomy boundary explicit: scientific choices that determine the meaning and scope of the calculation are agreed before resources are committed, whereas operational choices within the approved scope are delegated to the system. Beyond this boundary, the Orchestrator assigns tasks to specialist agents, coordinates handoffs and responds to results from scientific gates without requiring routine human intervention.

Domain Agents handle molecular structure preparation, MD and conformational sampling, QC reference calculations, ML model training and evaluation, and reporting. Each critical specialist is paired with a Scientific Assurance Agent (SAA) that evaluates the artifact produced at that stage (Fig. 1b,c). The Domain Agent retains responsibility for creating and revising the artifact; the SAA evaluates the artifact itself rather than relying on the producer's narrative summary. The SAA can assign three outcomes for an artifact. A pass permits downstream use within the approved scope when all requirements are satisfied. An inconclusive outcome withholds scientific acceptance but permits authorized diagnostic, recovery, or data-generation steps. The unresolved limitation remains attached to the artifact for later agents and reviewers. A failed artifact enters a recovery path before any downstream usage. Depending on the domain-specific recovery policy, the SAA either requests a local revision from the paired Domain Agent or escalates directly to the Orchestrator. Revised artifacts return to the SAA for re-evaluation, and the SAA records the post-revision outcome in an assurance report. The Orchestrator uses that report and the task-wide context to determine whether to continue, revise, rerun, reroute or terminate the affected branch.

Together, these roles separate scientific evaluation from artifact generation and workflow coordination. The Domain Agent performs and revises the scientific work, the SAA controls acceptance, and the Orchestrator determines the workflow-level response. Successful software execution is therefore not sufficient by itself for an artifact to advance. Domain-specific recovery policies also limit unproductive repair loops by accounting for the computational cost and context of each task. The same architecture applies across data generation, model construction and dynamics even though the artifacts and acceptance criteria differ.

### 2.2 Traceable autonomous construction of an ethylene S1 potential

To show how these design principles operate in practice, we trace PULSES through a specific model-building task: training a PyTorch Geometric SchNet model [19,20] for the ethylene $S_1$ potential-energy surface (Fig. 2). The prompt specifies the target state, model family and software framework because excited-state model construction requires scientific choices beyond those in a routine ground-state calculation. Given the cost of reference-data generation and the diversity of excited-state problems, these choices are agreed with the expert before execution. The Planner Agent analyzed the request, retrieved the relevant tools and skills, and proposed four steps: generate conformers by molecular dynamics, compute $S_1$ reference energies, train and optimize SchNet, and deliver the trained model. The user could accept the plan, request changes or answer clarification questions. The autonomous stage began only after plan approval.

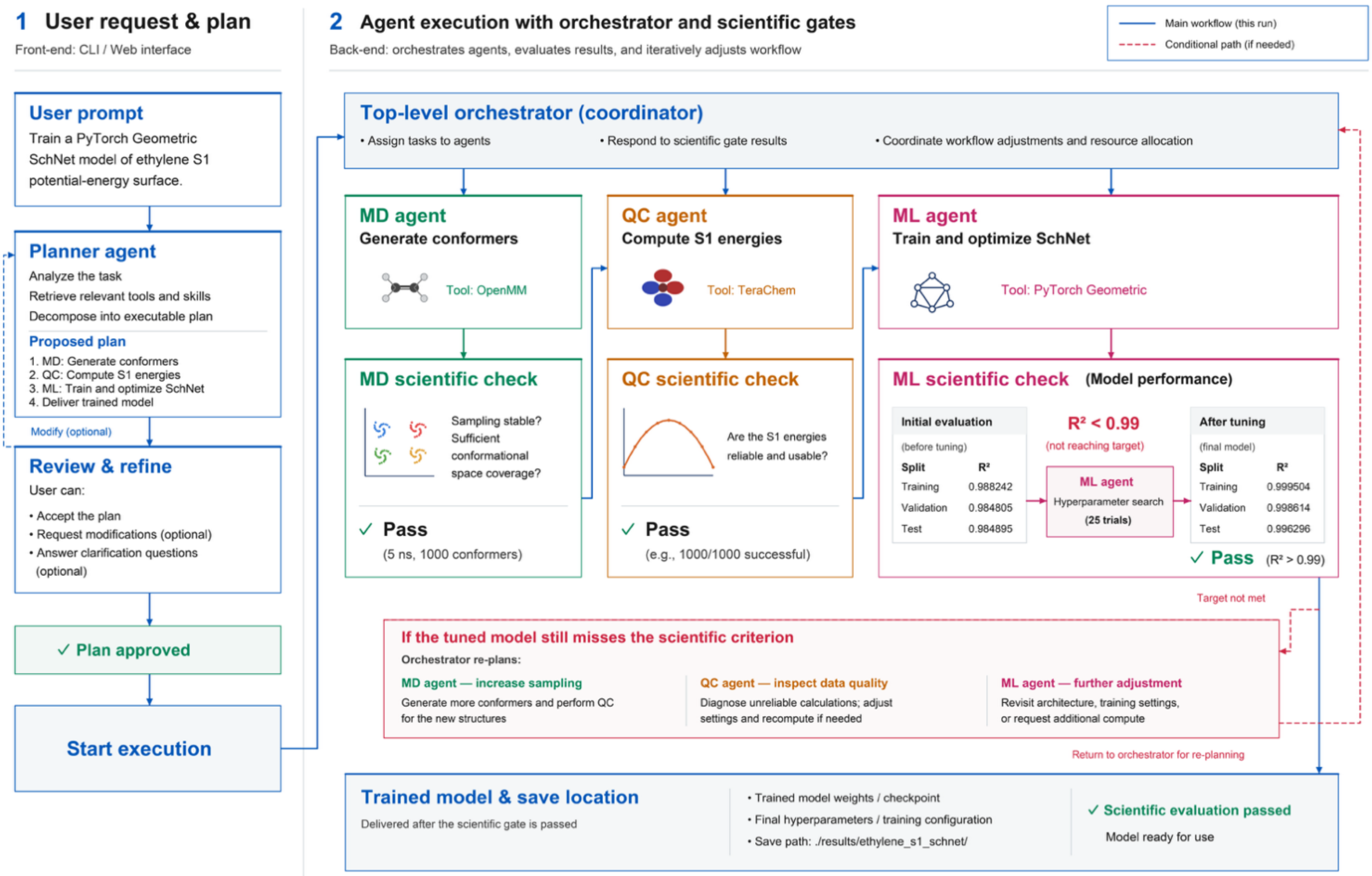

Fig. 2 Traceable autonomous construction and validation of an ethylene S1 SchNet model. Left, a user specifies the target potential-energy surface and model framework. The Planner Agent retrieves relevant capabilities, decomposes the task and presents a plan for optional revision and approval. Right, after approval, the Orchestrator assigns conformer generation, S1 energy calculations and model training to the MD, QC and ML agents. Scientific gates assess sampling stability and coverage, reference-data reliability and model performance. An initial model below the $R^2 > 0.99$ target triggers a 25-trial hyperparameter search; the tuned model exceeds the target across the training, validation and test sets. Solid blue lines show the executed path. Dashed red lines show the conditional replanning path that would increase sampling, inspect or recompute labels, or further revise the ML procedure if the final model failed the gate. The model and its training metadata are delivered only after the acceptance criterion is satisfied.

After approval, the Orchestrator assigned the sampling, electronic-structure and model-training stages to the MD, QC and ML agents, respectively (Fig. 2). The MD Agent used OpenMM [21] to generate 1,000 conformers from a 5 ns trajectory. Its scientific gate checked trajectory stability and coverage of conformational space before releasing the ensemble. The QC Agent then used TeraChem [22] to compute the S1 energies and passed the resulting dataset through a gate that assessed whether the calculations were complete, reliable and usable for training. Only the checked configuration ensemble and reference dataset were passed to the ML stage. Stage-specific checklists and recovery procedures are provided in Supplementary Note 5, with ethylene sampling, QC and model specifications in Supplementary Note 6.

Because the initial SchNet model fell slightly short of the predefined $R^2 > 0.99$ criterion across the three data splits, the ML Agent performed a 25-trial hyperparameter search and delivered a tuned model that exceeded the criterion on the training, validation and test sets, together with its checkpoint, hyperparameters and training configuration.

The solid path in Fig. 2 records the actions taken in this run, whereas the dashed path specifies the conditional response that would have been activated had tuning failed. Under that condition, the Orchestrator can expand conformational sampling, ask the QC Agent to diagnose unreliable labels and recompute them, or direct the ML Agent to revisit the architecture, training settings or resource allocation. Because the tuned model passed, these additional branches were not executed. The workflow therefore exposes both the executed provenance and the policy governing a failed gate. This example illustrates the capability targeted by PULSES: the system does not merely invoke an existing quantum-chemistry or dynamics tool, but constructs a new system-specific physics-inspired machine-learning model, evaluates it against a scientific acceptance criterion and delivers it for subsequent simulation.

### 2.3 General performance across excited-state-related tasks

We evaluated PULSES across 15 benchmark cases along three complementary dimensions: scientific task complexity, demonstrated scientific capabilities, and the accuracy and reproducibility of scientific outputs (Table 1). Existing benchmarks for large language model (LLM)-based scientific agents often emphasize final-answer accuracy, task completion, or performance on individual tasks [12,14]. These measures provide limited insight into the scientific demands of a task or an agent's ability to connect validated intermediate results into a complete workflow. This distinction is especially important in excited-state chemistry. Unlike many benchmarked ground-state calculations that terminate after a single electronic-structure task, an excited-state objective can require coordinated conformational sampling, ground- and excited-state electronic-structure calculations, state validation, model construction and dynamics. Our evaluation therefore examines both the validity of individual outputs and the dependencies connecting successive scientific stages in each requested workflow.

We designed the Scientific Task Complexity Index (STCI) to quantify the scientific complexity of each requested task. STCI combines workflow depth, integration across scientific domains, scientific constraints, potential silent-error checkpoints, and required refinement into a score from 0 to 100, independently of execution outcome. To examine how complexity changes with task scope, we included related problems at different workflow stages. The benzene π-stacking cases extend from grid construction and calculation submission to a complete interaction-energy surface, while the ethylene cases extend from molecular dynamics sampling and conformer selection to quantum-chemical labelling and machine-learning model development.

We used Verified Breadth–Depth (VBD) to evaluate the breadth and depth of the scientific capabilities demonstrated during execution. Reported as (B, D), breadth (B) counts distinct capability classes supported by validated outputs, and depth (D) counts the distinct capability classes connected along the longest validated sequence starting from the input. Each step must use the preceding step's output, with recorded evidence confirming that connection. Repeated operations within the same class do not increase the score. Together, STCI and VBD distinguish the complexity of the requested task from the capabilities demonstrated within that scope. Detailed scoring rubrics and validation procedures are provided in Supplementary Note 1.

We assessed accuracy and reproducibility by comparing agent-generated outputs with reference results obtained manually without agent assistance. For the repeated-run benchmarks, each task was executed three times using the same fixed prompt. Prompts were refined during benchmark development to resolve ambiguity and remained unchanged during evaluation. For tasks producing a single numerical result, accuracy was reported as the mean absolute error (MAE) across the three executions relative to the reference value. For workflows producing scans, spectra, datasets, models or trajectories, evaluation considered agreement with both the required computational stages and the reference scientific outputs. A task was labelled Pass (3/3) when all three executions completed every required stage and no discrepancy changed the scientific conclusion or invalidated a required workflow stage. Computationally intensive tasks were evaluated in a single run and reported separately as Validated (n = 1) when they met the corresponding scientific validation criteria. This designation indicates validation of the observed execution; it does not establish reproducibility across runs.

**Table 1. Scientific task complexity, verified capabilities and validation outcomes across multiagent workflow benchmarks.**

| Test | Scientific Task | Key Scientific Challenges | STCI | VBD | Accuracy |
|---|---|---|---|---|---|
| 1. Benzene (single-point energy energy) | B3LYP/6-31G* single-point energy | Geometry, electronic state, method/basis, convergence, energy units. | 25.0 | (2,2) | MAE($E_{elec}$) = 2.1 × $10^{-7}$ $E_h$ |
| 2. Benzene (ionization potential) | ΔSCF vertical IP | Neutral/cation states, restricted/unrestricted treatment, convergence, energy difference and units. | 35.8 | (4,4) | MAE (IP) = 1.8 × $10^{-5}$ eV |
| 3. Benzene (electron affinity) | ΔSCF EAs with a diffuse-basis comparison | Anion state, diffuse basis, SCF stability, EA sign and units. | 46.7 | (4,4) | MAE (EA) = 6.7 × $10^{-6}$ eV (def2-SVP) and 3.6 × $10^{-6}$ eV (def2-SVPD) |
| 4. Benzene ( $\omega$ tuning) | Coarse-to-fine range-separation parameter tuning | ω settings, HOMO energy, IP objective, refinement and optimum selection. | 57.5 | (4,4) | MAE ($\omega$) = 0.00 $au^{-1}$ |
| 5. Perylene (TDDFT spectroscopy) | Gas/neq-LR-PCM TDDFT spectra | Excited-state count, neq-LR-PCM, oscillator strengths, energy conversion, broadening. | 50.0 | (4,4) | Pass (3/3) |
| 6. Biphenyl (MD) | Solvated MD and torsional distribution analysis | Force-field setup, solvent box, ensemble/staging, trajectory integrity, torsion definition. | 54.2 | (3,3) | Pass (3/3) |
| 7. Benzene π-stacking (submission) | Automated construction and submission of a π-stacking conformation scan along the XY plane with 121 grid points. | Dimer placement, grid completeness, electronic state, QC inputs, geometry-job mapping. | 37.5 | (2,2) | Pass (3/3) |
| 8. Benzene π-stacking (surface) | Construction and analysis of a complete π-stacking conformation scan along the XY plane with 121 grid points. | Grid placement,, monomer reference, batch completeness, interaction-energy calculation, geometry-energy mapping. | 57.5 | (4,4) | Pass (3/3) |
| 9. Anthracene exciton-model scan | 2D π-stacking exciton-energy and coupling surfaces | Input geometry, dimer displacement, exciton-model setup, wPBEh/6-31G*, RESP/ESP settings. | 62.5 | (4,4) | Pass (3/3) |
| 10. Ethylene MD (sampling) | MD sampling and 1,000-conformer data handoff | Gas-box setup, 400 K NVT, monomer selection, sample coverage, schema and units. | 51.7 | (2,2) | Pass (3/3) |
| 11. Ethylene MD-QC-ML | MD sampling, TDDFT labeling, and SchNet training | MD sampling, S1–S5 labels, data integrity, train/test separation, model validation. | 91.7 | (4,4) | Pass (3/3) |
| 12. Perylene dimer excitation-energy scan | Monomer optimization and cofacial dimer S1–S5 distance scan | Monomer geometry, dimer separation/alignment, LE/CT set up, ESP charge fitting, excitation-energy extraction, scan completeness. | 62.5 | (5,4) | Pass (3/3) |
| 13. Perylene exciton model | ML prediction of the 4×4 Frenkel exciton Hamiltonian | Dimer coverage, LE/CT mapping, QC labels, data splits, model validation, uncertainty and targeted resampling. | 100.0 | (4,4) | Validated (n=1) |
| 14. Biphenyl Excited-state relaxation in Solvents | Training and deploying an excited-state MACE potential to simulate biphenyl torsional relaxation in nine solvents | Coverage of torsional configurations, accurate $S_1$ energy and force predictions, consistent coupling of MACE with the solvent force field, stable dynamics, reliable estimation of relaxation. | 85.0 | (7,6) | Validated (n=1) |
| 15. HBQ Excited-state intramolecular proton transfer | Sampling, $S_0$/$S_1$ QC, ML potentials, and excited-state dynamics | Proton-transfer coordinate, state/gradient labels, force units, model selection, dynamics stability, crossing criterion. | 95.0 | (7,6) | Validated (n=1) |

Table 1 highlights the range of scientific demands PULSES can handle, with STCI scores ranging from 25.0 to 100.0. The simpler cases test the correct specification and execution of individual electronic-structure calculations, while the more complex cases require coordination of sampling, quantum-chemical labelling, dataset preparation, ML model development and downstream simulation. This progression is reflected in VBD: the benzene single-point calculation scores (2,2), whereas the complete biphenyl training–relaxation workflow scores (7,6), linking configuration generation to dynamics with learned potentials and analysis of a defined relaxation observable. The benchmark therefore assesses both individual computational operations and the validated dependencies that connect them into scientific workflows. The following sections examine the three complex cases in detail to illustrate how the framework coordinates these stages and addresses the scientific challenges encountered during execution.

### 2.4 Autonomous construction and evaluation of a perylene exciton model

We next asked PULSES to construct a machine-learned exciton Hamiltonian for perylene dimers using the framework of Ren et al. (2025) [23]. This model predicts the energies of local-excitation (LE) and charge-transfer (CT) states and the couplings between them from molecular geometry. Diagonalizing the Hamiltonian then yields the dimer's adiabatic excitation energies and state compositions. We retained one LE state per monomer and two oppositely directed CT states, giving a four-state Hamiltonian with nine independent learned elements [24] (Fig. 3a). The expert-defined task specified the Hamiltonian, electronic-structure method, sampling range and evaluation criteria. The agents generated the calculation inputs, reference labels and trained models.

The workflow separated inexpensive candidate generation from costly quantum-chemical labelling. The MD Agent prepared a pool of 13,000 dimer configurations combining MD-derived structures with controlled rigid-body configurations (Supplementary Note 7, Section 7.2). Intermolecular separation was an explicit consideration in both dataset construction and model evaluation: a model that performs well overall may still predict poorly within a particular separation range. We therefore grouped configurations into intervals of $d_{AB}$, the closest carbon-carbon contact distance between the monomers, and assessed accuracy separately within each interval. The initial training, validation and test sets contained 1,600, 400 and 1,000 dimers, respectively.

For the selected configurations, the QC Agent calculated reference Hamiltonians and applied phase correction to establish consistent coupling signs. The ML Agent then trained and optimized neural-network ensembles for the nine Hamiltonian elements. Acceptance checks assessed the accuracy of each matrix element and of the four excitation energies obtained by diagonalization, with an initial target of $R^2 \geq 0.95$ both for the overall set and within every distance interval. After initial training and hyperparameter optimization, interval-specific checks remained below target across the sampled distance range. The Orchestrator consequently requested additional QC labelling from the existing reserve pool (Fig. 3b). One expansion round added 1,300 dimers, bringing the training set to 2,900 (Fig. 3c).

After retraining, all nine Hamiltonian elements achieved overall test-set $R^2 \geq 0.990$, with root-mean-square errors (RMSEs) of 0.484–13.8 meV. The four excitation energies had RMSEs of 2.50–13.7 meV. Despite the overall agreement, some distance-specific checks nevertheless remained below the $R^2$ target. The Orchestrator examined absolute prediction errors alongside the variation in the reference values. Most remaining failures involved intervals with very little reference variation, where even small absolute errors can produce low or negative $R^2$ because the metric normalizes squared errors by the reference variance. All remaining failed checks had RMSEs below 43 meV. On the basis of these diagnostics, the Orchestrator stopped further expansion while recording that the strict interval criterion remained unmet. This outcome represents termination with residual criterion failures, rather than satisfaction of all predefined accuracy requirements. Supplementary Note 7, Section 7.4 details the diagnostics and stopping decision.

Because validation and test diagnostics had informed data expansion and stopping, these results describe the adaptive development campaign rather than an evaluation on an unseen set. To assess performance beyond the development data, we subsequently evaluated the model performance on 100 unseen dimer configurations sampled from the production MD trajectory. RMSEs ranged from 1.48 to 66.21 meV for the nine learned Hamiltonian elements and from 25.17 to 64.20 meV for the four excitation energies obtained by diagonalization (Fig. 3d).

Finally, the learned Hamiltonian was used to obtain an ensemble-averaged spectrum for the same 100 production dimers and compared with QC spectra evaluated on the same configurations (Fig. 3e). Diagonalization supplied excitation energies and eigenvectors, which were combined with QC diabatic transition dipoles to obtain adiabatic transition dipoles and oscillator strengths. Individual spectra were broadened using Gaussians with a full width at half maximum of 0.060 eV and averaged with equal weights.

Across the four excited states of each dimer, the MAE relative to QC exciton-model excitation energies was 36.7 meV. The ML monomer energies, constructed using dimer model-predicted local-excitation energies, had an MAE of 23.2 meV relative to the QC excitation energies of the 200 constituent monomer geometries. This application completed the workflow from candidate generation and reference calculations through model construction, adaptive dataset expansion, and deployment for a spectroscopic calculation.

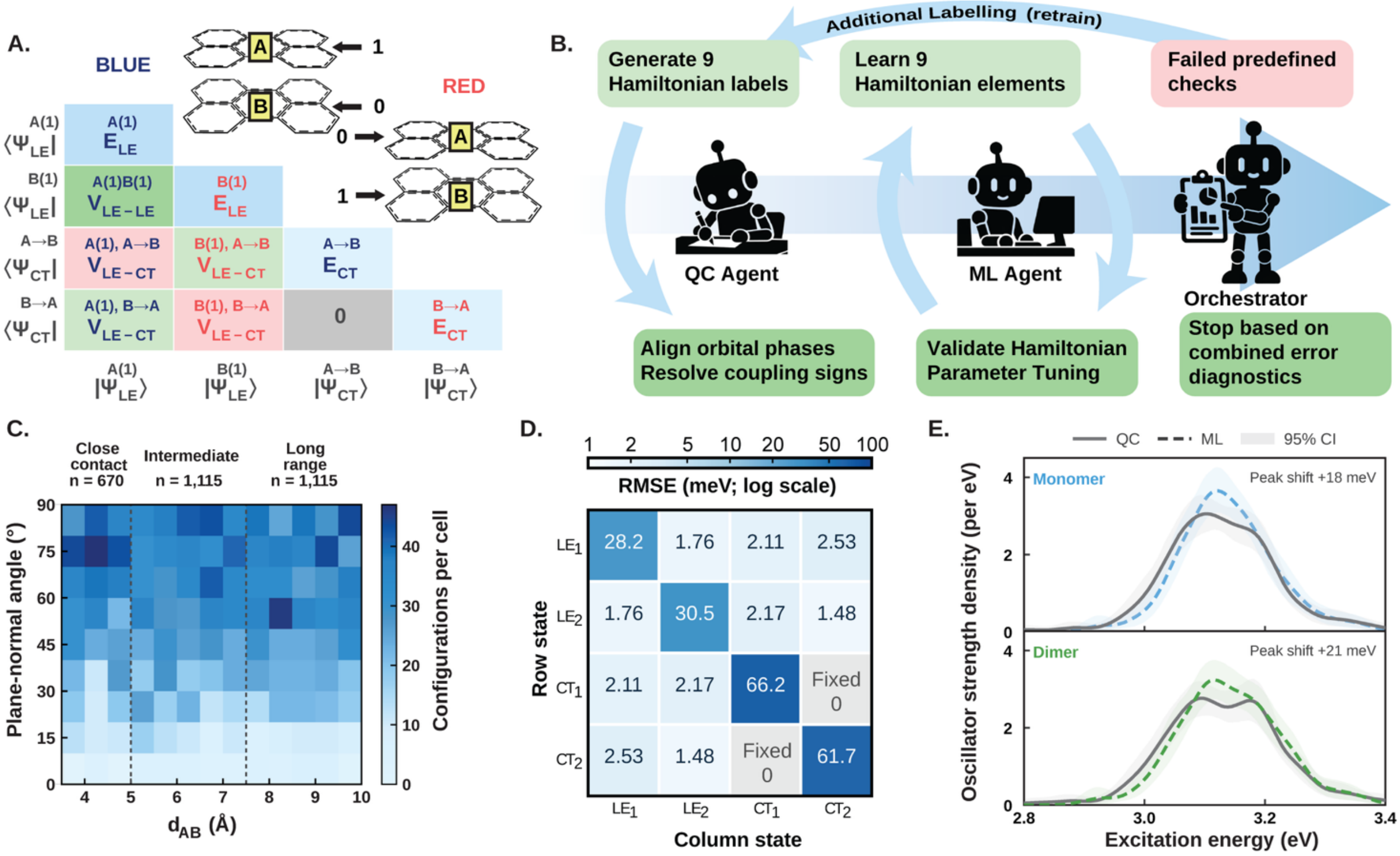

**Figure 3. Multi-agent construction and validation of a perylene exciton model.** **a** Four-state diabatic basis comprising two local-excitation (LE) and two bidirectional charge-transfer (CT) states. Nine independent nonzero Hamiltonian elements are learned, whereas the CT–CT coupling is fixed at zero. **b** Autonomous workflow linking QC labelling and phase alignment, model training and evaluation, additional labelling and retraining after failed checks, and the Orchestrator's stopping decision based on combined error diagnostics, with residual interval-criterion failures recorded. **c** Distribution of the final 2,900-dimer training set over minimum intermolecular C–C distance (dAB) and plane-normal angle. **d** Elementwise RMSEs in meV for ML predictions on 100 unseen production-MD dimer configurations, arranged in the symmetric $4 \times 4$ Hamiltonian. **e** Gaussian-broadened production-ensemble spectra comparing the QC

reference and ML predictions for the monomer (top) and dimer (bottom). Curves show the mean oscillator-strength density after Gaussian broadening (FWHM, 0.060 eV). Shaded regions indicate sampling uncertainty in the ensemble-mean spectra, estimated as pointwise 95% bootstrap intervals from 2,000 resamples of five-configuration blocks. QC and ML results were resampled together to preserve pairing.

### 2.5 End-to-end execution of complex workflows: Biphenyl excited-state relaxation in solvents

Spectroscopic studies of biphenyl and its derivatives indicate a twisted S0 ground-state geometry and a near-planar S1 excited-state geometry [25], and a previous experimental study reported solvent-dependent structural relaxation for a para-substituted biphenyl derivative [18]. This process provides a stringent yet tractable test of autonomous model building: the structural relaxation can be followed on the $S_1$ potential-energy surface without modeling nonradiative decay or surface crossing, whereas resolving its solvent dependence requires an ensemble of explicit-solvent trajectories that would be costly with direct ab initio dynamics. We therefore used PULSES to construct and deploy system-specific excited-state MLIP for unsubstituted biphenyl. We chose the parent molecule rather than the larger long-chain derivative used experimentally to limit the cost of training-data generation, so the experiment serves as a qualitative rather than molecule-specific benchmark.

The MD Agent first parameterized biphenyl with the GAFF2 force field using AutoSolvate [26] and performed unbiased ground-state conformational sampling. After identifying insufficient coverage of the near-planar region associated with the S1 minimum, it initiated PLUMED metadynamics along the inter-ring torsion [27]. The resulting 2,000 frames spanned the full 0–180° torsional range. The pipeline retained 1,000 geometrically diverse structures for QC labeling with S0 and S1 energies and forces (Fig. 4b) and divided them into 800 training, 100 validation and 100 test structures. When initial training did not meet the prespecified force-accuracy criterion, the ML Agent reconfigured and retrained the MACE models [28]. The final models reproduced test-set forces accurately on both surfaces ($R^2$ = 0.9991 and 0.9721 for S0 and S1, respectively), with larger errors for S1 than S0 for both force and energy predictions (Fig. 4c).

Consistent with the spectroscopic picture, the calculated torsional profiles favor a twisted S0 geometry and a near-planar S1 geometry (Fig. 4a).

The trained S0 and S1 MACE potentials were deployed for the biphenyl solute in explicit-solvent simulations. We used OpenMM [29] to couple the solute fitted MACE model to a classical solvent environment described by GAFF2 [30]. After 20 ps of S0-state NPT equilibration at 298 K and 1 bar, each solvated system was propagated for a further 192 ps, with configurations saved every 4 ps. The resulting 48 configurations per solvent served as initial conditions for vertical excitation, modeled by switching the solute potential from S0 to S1 while retaining the nuclear coordinates, velocities and box vectors. Each trajectory was then propagated on S1 for 100 ps with a 0.5-fs time step, yielding 48 trajectories for each of the nine solvents (432 total). The angle θ between the phenyl-ring planes was recorded every 10 fs. To quantify torsional relaxation across trajectories with different starting and final angles, we defined the relaxation time $\tau_{1/e}$ as the first recorded time at which θ crossed the trajectory-specific 1/e threshold,

$$\theta_{\mathrm{thresh}}=\theta_{\mathrm{late}}+\frac{(\theta_{\mathrm{start}}-\theta_{\mathrm{late}})}{e},$$

Here, the starting angle is the first recorded angle and the late-time angle is the median over the final 10% of the trajectory. The threshold corresponds to completion of 1 - 1/e (approximately 63%) of the angular relaxation from the starting to the late-time angle. Using the late-time median reduces sensitivity to fluctuations around the relaxed geometry and provides a consistent trajectory-specific timescale. Figure 4d illustrates the early torsional evolution in methanol and marks its mean individual-trajectory threshold time of 0.389 ps. Across the six aprotic solvents, mean threshold times broadly increased with viscosity, ranging from 0.254 ps in acetonitrile to 3.519 ps in n-hexadecane (Fig. 4e). Methanol, ethanol and 1-butanol yielded means of 0.389, 0.233 and 0.501 ps, respectively, below the aprotic-solvent trend. These observations qualitatively parallel the experimental findings for a para-substituted biphenyl derivative [18].

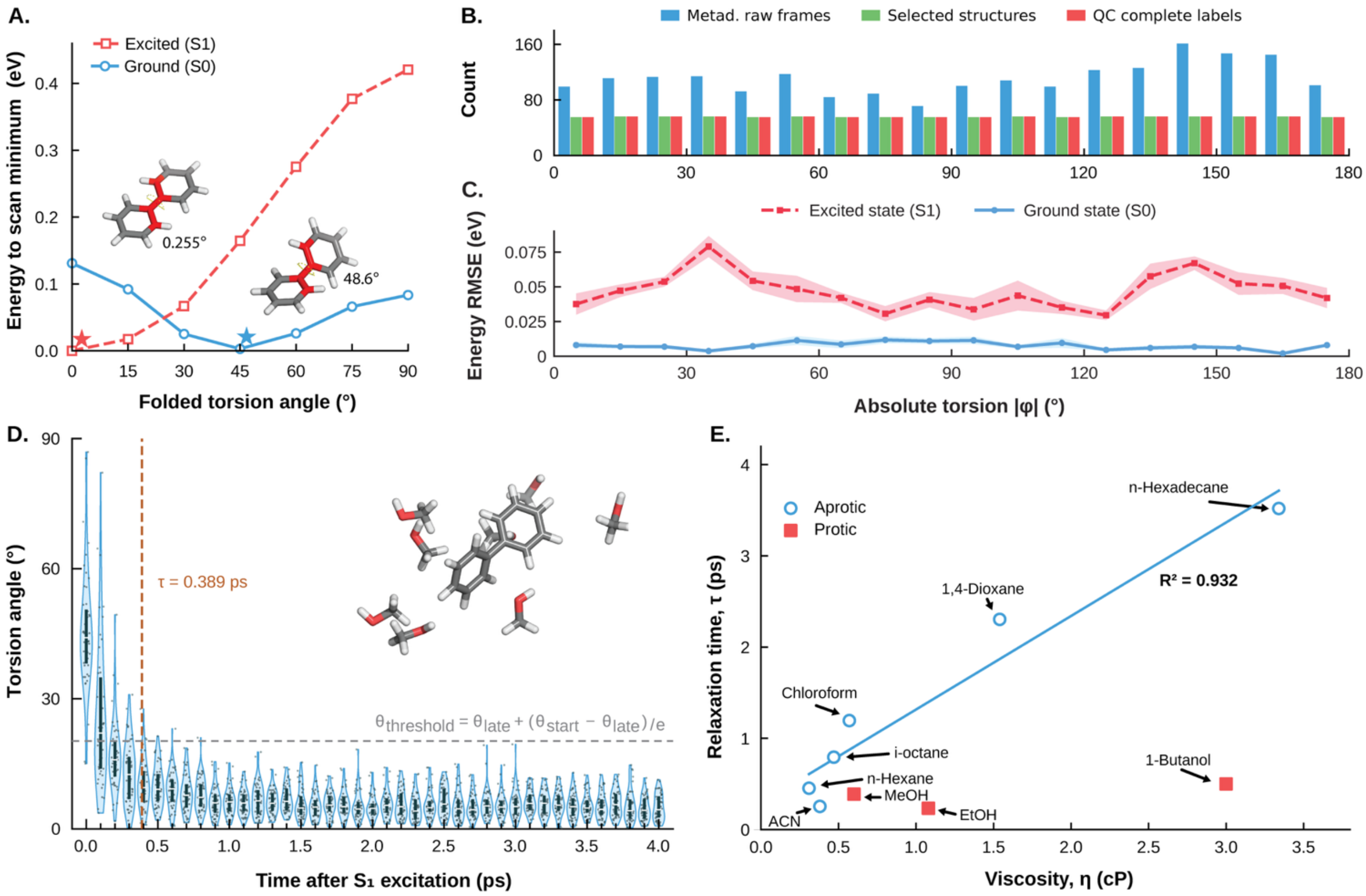

**Figure 4. Agent-directed biphenyl excited-state relaxation in nine solvents. a** Ground-state and excited-state energy profiles predicted by agent-trained MACE models as a function of the inter-ring dihedral angle. **b** Dihedral angle distributions of structures generated by well-tempered meta-dynamics and the subset selected for MACE model training. **c** Energy prediction errors of the $S_0$ and $S_1$ machine-learned potentials relative to quantum-chemical reference calculations, plotted as a function of the dihedral angle for the same 100 test geometries. **d** Distributions of biphenyl torsion angles across 48 trajectories in methanol following $S_1$ excitation. **e** Torsional threshold crossing times $\tau_{1/e}$ (see text) in aprotic and protic solvents plotted against solvent viscosity; the line shows a linear fit for the aprotic solvents.

### 2.6 End-to-end execution of complex workflows: Gas-phase excited-state proton transfer in HBQ

To evaluate PULSES on an end-to-end reactive excited-state workflow, we tasked it with estimating the gas-phase excited-state intramolecular proton-transfer (ESIPT) timescale of 10-hydroxybenzo[h]quinoline (HBQ). HBQ provides a stringent yet tractable benchmark. Proton transfer breaks and forms covalent bonds, so conventional fixed-topology force fields are inadequate and the dynamics require an ab initio or learned reactive potential. At the same time, previous ML simulations of HBQ ESIPT have treated the reaction on S1, allowing the dynamics to be modeled on a single excited-state surface rather than with a full nonadiabatic treatment [31]. Raucci et al. reported an essentially barrierless transfer within approximately 50 fs, accompanied by an emission-energy redshift of approximately 1 eV [31].

The central challenge was constructing state-specific potential-energy surfaces that adequately represented the reactive region. The Orchestrator first directed the Structure Agent to verify the tautomeric identity of the PubChem-derived geometry [32] and define the proton-transfer coordinate $\delta = r(O–H) - r(N–H)$, with negative and positive values denoting enol-like and keto-like structures, respectively (Fig. 5a).

The MD Agent used GFN-xTB [33] dynamics combined with PLUMED [34] well-tempered metadynamics to generate configurations spanning both tautomeric basins and the intervening region. After filtering and deduplication, the QC Agent labelled 1,000 geometries, comprising 350 enol-like, 300 intermediate-region, and 350 keto-like configurations (Fig. 5b). The ML Agent trained separate MACE potentials for the S0 and S1 surfaces using an 800/100/100 training, validation, and test split. The test partition was excluded from fitting but was consulted during S1 refinement and final workflow acceptance. An initial S1 candidate passed the validation force criterion but missed the test force-components $R^2$ threshold, which prompted further model tuning. In the end, the S0 and S1 models achieved force-component $R^2$ values of 0.9962 and 0.9905, with RMSEs of 0.077 and 0.123 eV $Å^{-1}$, respectively. In the post-hoc evaluation, predicted energies were corrected by subtracting state-specific offsets estimated from the validation partition (−0.072 eV for S0 and +0.123 eV for S1). This yielded the corrected test-set energy MAEs of 0.0246 eV for S0 and 0.0241 eV for S1. (Fig. 5c). The test geometries were excluded from parameter fitting, but test performance informed S1 model refinement. The reported S1 test metrics therefore characterize adaptive model development rather than performance on an unseen final test set.

The S0 MACE model was then used to propagate a continuous 1-ns trajectory at 300 K under Langevin NVT dynamics with a 0.5-fs time step. One hundred initial conditions, comprising coordinates and velocities, were selected at 10-ps intervals. Vertical excitation was modeled by switching to the S1 MACE potential while retaining these coordinates and velocities, followed by 1 ps of NVE dynamics with the same time step. Proton transfer was defined as the first crossing of the proton-transfer coordinate from negative to nonnegative values, which occurred within 100 fs in 75 of the 100 trajectories. The mean individual first-passage time was 58.25 fs in this fast-transfer subset. The remaining 25 trajectories formed a slower tail, resulting in the ensemble-averaged proton-transfer coordinate crossing zero at 67.47 fs (Fig. 5d).

Reproducing this trajectory ensemble with direct on-the-fly electronic-structure dynamics would require approximately 2.2 million energy-and-gradient evaluations for the 2,000,000 S0 and 200,000 S1 integration steps. Raucci et al. [31] substantially reduced this cost and achieved higher force accuracy through an expert-designed active-learning workflow involving a domain-specific reaction coordinate, iterative OPES [35] sampling, committee-based uncertainty estimation and repeated model refinement. PULSES addresses a different objective: autonomous construction of a fit-for-purpose model under predefined scientific checks, rather than reproduction of the accuracy of a bespoke active-learning campaign. It completed structure preparation, reactive sampling, QC labelling, training of the S0 and S1 potentials and the final 100-trajectory ensemble in 17.2 h of recorded wall time. Although its potentials were less accurate than the expert-designed model, they recovered the expected enol-to-keto pathway and placed the transfer in the correct ultrafast regime. The workflow is therefore best viewed as an autonomous first-pass mechanistic screen that can identify a viable pathway and timescale before expert-guided refinement when quantitative accuracy is required (Fig. 5d).

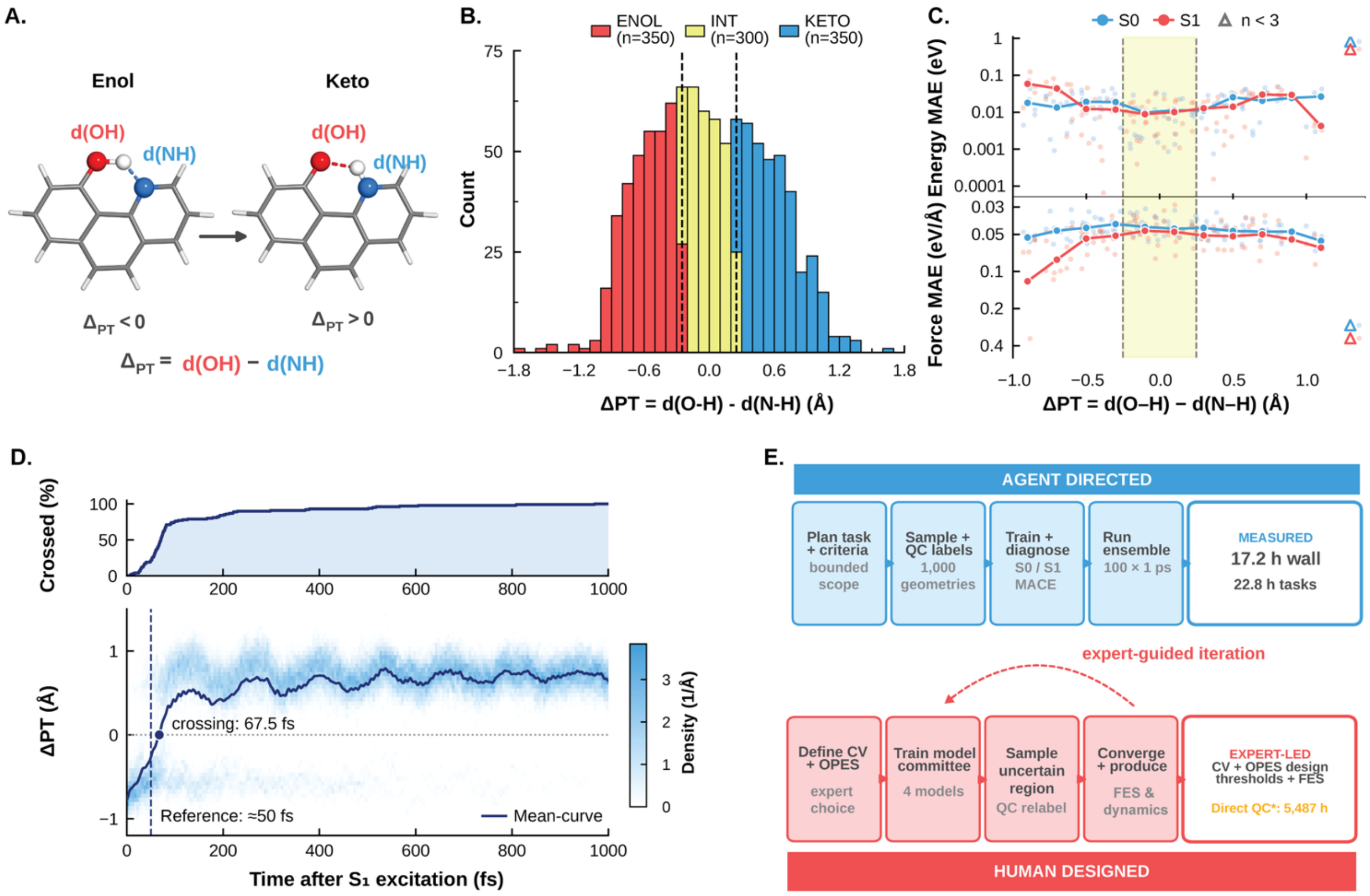

**Figure 5. Agent-directed modelling of excited-state proton transfer in HBQ. a** Enol and keto forms of HBQ and definition of the proton-transfer coordinate, $\Delta_{PT} = d(O-H) - d(N-H)$. **b** Coverage of the enol-like, intermediate and keto-like regions in the quantum-chemically labelled configuration set. **c** Energy and force prediction errors of the $S_0$ and $S_1$ MACE models across the proton-transfer coordinate; the shaded region denotes intermediate configurations. **d** Cumulative proton-transfer probability and time-resolved distribution of the trajectory ensemble with the mean trajectory and reference timescale indicated. **e** Comparison of the PULSES workflow with the expert-designed active-learning workflow reported by Raucci (ref. 31). The value marked "Direct QC*" (5,487 h) is our estimated serial-equivalent cost of propagating the same trajectory workload as in this study using direct QC dynamics, calculated from integration-step counts and mean QC gradient timings. It is not the computational cost of Raucci's workflow. The PULSES timings denote controller wall time (17.2 h) and summed recorded task durations (22.8 h), which include overlapping tasks. These quantities do not constitute a measured end-to-end speedup comparison.

## 3. Discussion

PULSES advances agentic computational chemistry from operating established software to constructing, validating and deploying fit-for-purpose scientific ML models. Across the three applications, the system addressed distinct physical regimes using models tailored to the corresponding scientific objectives. It constructed a multistate excitonic Hamiltonian to predict collective excited states and spectra in perylene assemblies; trained state-specific interatomic potentials to simulate nonreactive torsional relaxation of biphenyl through long-timescale trajectory ensembles in explicit solvents; and developed reactive excited-state potentials to capture ultrafast intramolecular proton transfer in gas-phase HBQ. The latter two applications employ the same general class of machine-learning potential but place fundamentally different demands on model construction and deployment: one describes solvent-dependent conformational relaxation without bond rearrangement, whereas the other must represent bond breaking and formation along a reactive coordinate. The main contribution is therefore not the automation of any individual quantum-chemistry calculation, but an architecture that coordinates data generation, electronic-structure calculations, model construction and downstream simulation across scientific domains. PULSES implements expert-bounded rather than unrestricted autonomy: consequential choices concerning the scientific objective, physical approximations and computational methods are agreed upon before execution, whereas the resource-intensive operational workflow is delegated to the agent system.

Scientific Assurance Agents provide a second level of control after autonomous execution begins. By evaluating scientific artifacts before they propagate downstream, these agents make validation outcomes and unresolved uncertainty explicit and traceable. This separation of artifact generation from artifact acceptance is particularly important for complex workflows, in which a technically successful calculation can nevertheless contain an incorrect state assignment, inadequate sampling, inconsistent units or an unsuitable dynamics approximation. The assurance layer does not guarantee scientific correctness, however. Domain Agents and their paired assurance agents may share the same underlying language model and can therefore exhibit correlated reasoning errors. Moreover, the present study does not include a systematic ablation comparing PULSES with and without the assurance layer. Future evaluations should quantify error detection, silent-failure escape rates and recovery costs, and should examine whether heterogeneous language models or deterministic scientific validators further improve reliability.

The current implementation focuses on dynamics propagated on separately modelled adiabatic electronic-state surfaces, providing a well-defined foundation for extending PULSES to more general photochemical processes. A major next step is to construct models of nonadiabatic couplings and to execute surface-hopping [36] or related nonadiabatic dynamics protocols [37]. Such an extension will require coordinated learning and validation of multistate energies, gradients and nonadiabatic or diabatic couplings, together with reliable state tracking and phase consistency near degeneracies [38]. PULSES also provides the architectural components needed for active learning [39]: when a model fails an assurance gate, the system can already request additional sampling, quantum-chemical labelling and model revision. Connecting these operations through uncertainty-driven triggers [40] would establish a fully closed learning cycle in which coverage gaps or configurations encountered during dynamics initiate targeted data generation, retraining and revalidation. These developments would extend PULSES from autonomous construction of system-specific adiabatic models to continuously improving models for multistate photochemical dynamics.

Controlled comparisons across models, repeated executions and matched computational budgets are needed to distinguish model effects from architectural effects. PULSES should therefore be viewed not as a substitute for scientific expertise, but as infrastructure that converts an expert-approved scientific strategy into a traceable, checked and computationally scalable execution.

## 4. Method

### 4.1 PULSES architecture and execution

As shown in Figure 1, PULSES organizes scientific workflows around a planner, an orchestrator, domain agents, and SAAs. The planner converts research requests into saved execution plans, which the orchestrator uses to delegate tasks, coordinate dependencies, and manage revisions. Domain agents perform molecular structure preparation, MD and conformational sampling, QC calculations, ML training and evaluation, and scientific reporting. SAAs inspect workflows, evaluate outputs, diagnose failures, and conduct scientific review through task-specific pathways.

Agents exchange instructions and artifact references through a shared filesystem containing scripts, datasets, workflow manifests, calculation-status tables, and provenance records. The filesystem also stores agent skill files in the form of Skills.md, which provide task-specific instructions for workflow execution (Supplementary Note 3). Plan snapshots, runtime profiles, and audit histories support traceability throughout execution. When the calculation jobs finish, scheduler continuations restore saved conversation and task state. Scientific execution combines callable tools, scripts generated through the shell, and an AutoSolvate MCP service with managed Python environments and native scientific executables. Slurm schedules CPU and GPU calculations, while agents inspect scheduler states, logs, and parsed outputs to determine subsequent actions.

**4.2 Planning and execution contracts**

In interactive workflows, an expert user and the planner work together to translate a research request into an execution contract. The planner interprets the objective, proposes methods and ordered workflow stages, and identifies missing inputs and assumptions. User feedback revises the complete saved plan before approval and execution. The contract follows OMWAB: the scientific objective (O), approved methods (M), workflow structure (W), acceptance criteria (A), and resource and decision boundaries (B). Following user approval, the application freezes the plan as a hash-identified snapshot and passes it directly to the orchestrator without another planning turn.

Within the approved scope, agents implement scripts, allocate work, tune permitted parameters, and repair execution failures while preserving unaffected results. Changes to the scientific objective, prescribed methods, or acceptance criteria require user approval unless covered by an approved contingency. Terminating a run with unmet criteria does not establish scientific acceptance or revise those criteria; unresolved failures must remain explicit in the report. Missing information and work beyond authorized resource or decision boundaries will wait for user approval for continuation. Routine execution failures first return to the orchestrator for targeted repair. Unresolved cases are reported to the user with the decision required to continue execution. Details of plan approval, versioning and execution boundaries are provided in Supplementary Note 4.

**4.3 Scientific validation and revision**

PULSES combines execution with artifact-specific validation and targeted revision through domain-specific checkpoints. Deterministic checks assess file and manifest consistency, execution completion, provenance, and applicable numerical and structural criteria (Supplementary Note 5). SAAs assess whether the recorded methods and results support the approved objective. This assessment considers physical plausibility, methodological consistency, and unresolved limitations in the available evidence. PASS means that the available evidence satisfies the requirements for the specific claim, workflow stage, and intended use, permitting advancement to downstream tasks within the approved scope. INCONCLUSIVE means that the evidence is insufficient to accept or reject the artifact. It withholds scientific acceptance and permits only authorized diagnostic, recovery, or data-generation steps with the unresolved limitation retained. FAIL means that a requirement has been violated and prevents the downstream use that depends on the requirement until repair and successful re-evaluation, or termination of the affected branch.

The orchestrator routes findings from the SAAs to the responsible domain agent, which revises the affected inputs, implementation, or analysis before revalidation. Missing information and repairs beyond the approved scope return to the user. Artifacts with failed or unresolved blocking checks cannot pass downstream acceptance checks. Explicitly nonblocking warnings permit continued execution but remain visible in review and reporting. Before a large sweep, the MD or QC agent executes and evaluates one preliminary pilot entry. Approval of this entry supports high-throughput workflow reusing it without certifying individual production results. Local retry guards limit repeated attempts that reproduce the same failure.

### 4.4 Benchmark design and evaluation

PULSES was evaluated on 15 tasks spanning electronic-structure calculations, spectroscopy, molecular sampling, exciton modelling, model development, and excited-state dynamics. Evaluation separately characterized task complexity, demonstrated capabilities, and agreement with reference outcomes, which came from validated manual executions performed without agent assistance or from literature reports. We quantified task complexity using STCI and verified capability breadth and depth using VBD for all 15 tasks (Section 2.3).

In our benchmark, 12 tasks were executed three times using the same fixed prompt. Scalar outputs were compared with reference values using mean absolute error. Workflows were assessed for protocol adherence, completion, artifact validity, and agreement with reference scientific outcomes. Tasks meeting these criteria without human intervention in all three executions were considered passing. To limit token usage, three computationally intensive workflows were evaluated once. This designation reflects single-run validation and does not establish repeatability. Scoring definitions and task-level scores are provided in Supplementary Note 1 and Supplementary Tables 1–4; task prompts appear in Supplementary Note 2.

### 4.5 Ethylene excitation-energy prediction

For the task described in Results Section 2.2, a neutral ethylene structure from PubChem was parameterized using AutoSolvate, GAFF [41] and AM1-BCC charges. The resulting system comprised ten ethylene molecules in a periodic simulation box. After minimization and 50 ps of heating, OpenMM sampled the gas-phase system at 400 K for 5 ns under NVT conditions. The simulation used a 1 fs timestep and saved snapshots every 250 steps. From the 20,000 saved frames, 1,000 conformers were selected deterministically using evenly spaced indices. TeraChem calculated the five lowest singlet vertical excitation energies at the B3LYP/6-31G* level [42,43] using the Tamm–Dancoff approximation.

A seeded random split assigned 800 conformers to training, 100 to validation, and 100 to testing. SchNet was trained from scratch to predict $S_1$ excitation energies, $\Delta E_1(R) = E_{s1}(R) - E_{s0}(R)$, evaluated at a fixed nuclear geometry, with target values across all three subsets standardized using training-set statistics. A baseline model and 24 random-search trials used validation RMSE for checkpoints and model selection. The retained model used four interaction blocks, 256 hidden channels, 64 filters, 50 Gaussian basis functions, an 8 Å cutoff, and additive readout. Acceptance was satisfied with $R^2 > 0.99$ on both the validation and test partitions. The selected model achieved a test-set $R^2$ of 0.9963 and an RMSE of 10.1 meV. Further sampling, training and evaluation details are provided in Supplementary Note 6.

**4.6 Perylene exciton Hamiltonian learning**

Perylene was parameterized using GAFF and AM1-BCC [44] charges in a 50 Å cubic box containing 402 molecules. OpenMM performed minimization, heating, and 10 ns of production dynamics at 300 K. The candidate pool used for training contained 13,000 dimers assembled from controlled rigid-body configurations and heating-trajectory structures. Configurations were grouped into thirteen 0.5 Å intervals of minimum intermolecular C–C distance. The initial training, validation, and test sets contained 1,600, 400, and 1,000 dimers, respectively. The remaining configurations were reserved for uncertainty reporting and potential training set expansion.

TeraChem calculations at the ωPBEh/6-31G* level [45,46] supplied four diabatic energies and five couplings. Preprocessing applied phase correction, and each target was fitted with a three-member neural-network ensemble. Validation and test sets both informed data expansion. Following unsuccessful acceptance checks, one expansion selected 100 reserve dimers per distance interval from the remaining 8,700 dimer configurations for labeling and retraining, yielding 2,900 unique dimers. Monomer-swap augmentation was performed on the 2,900 dimers, producing 5,800 training examples. Candidate selection, evaluation criteria and error diagnostics are detailed in Supplementary Note 7; metric definitions and ensemble seeds are provided in Supplementary Note 10.

For the additional evaluation on unseen data, we sampled 100 frames at 40 ps intervals from the final 4 ns of the production trajectory, and randomly selected one dimer from each frame conditional on a minimum intermolecular C–C distance of 3.5–10 Å.

### 4.7 Biphenyl torsional relaxation in solution

Separate $S_0$ and $S_1$ MACE potentials were trained for neutral biphenyl ($C_{12}H_{10}$) using 1,000 isolated-solute configurations selected from well-tempered metadynamics [47] in explicit cyclohexane. C–H bond lengths were constrained during sampling to reduce the number of sampled degrees of freedom. Reference energies and gradients were calculated using TeraChem at the ωB97X-D3/6-31++G** level [48]. Torsion-stratified partitions contained 800 training, 100 validation and 100 test geometries.

The potentials were coupled to classical solvent force fields for simulations in n-hexane, acetonitrile, isooctane, chloroform, methanol, ethanol, 1,4-dioxane, 1-butanol and n-hexadecane. MACE described biphenyl intramolecular interactions, while classical force fields described solvent interactions and solute–solvent electrostatic and van der Waals terms. For each solvent, 48 $S_0$ snapshots separated by 4 ps initialized vertical switches to $S_1$ while preserving positions, velocities, and boundary conditions. Each $S_1$ trajectory was propagated for 100 ps at 298 K and 1 bar using a 0.5 fs timestep.

The relaxation estimator was $\tau_{1/e}$ estimated by measuring the first-time sampled trajectory crossing $\theta_{thresh}$ (see Section 2.5). Solvent-level estimates were arithmetic means of these crossing times calculated from the 48 trajectories. External dynamic viscosity values provided the comparison axis across the solvents.  Training settings, trajectory protocols, torsion-resolved error analysis and the viscosity regression are detailed in Supplementary Note 8.

### 4.8 Excited-state proton transfer in HBQ

Gas-phase excited-state proton transfer in neutral HBQ was simulated using separate $S_0$ and $S_1$ MACE potentials. An enol reference structure was constructed while preserving the atom mapping. The proton-transfer coordinate was defined as $\delta = r(O\text{–}H) - r(N\text{–}H)$, with negative and positive values denoting O-bound and N-bound configurations, respectively.

TeraChem GFN-xTB dynamics biased by PLUMED well-tempered metadynamics generated enol, neutral, and keto configurations. Geometry filtering, within-bin RMSD deduplication, and coordinate-stratified selection retained 1,000 geometries. Each geometry received paired $S_0$ and $S_1$ energy and gradient labels at the ωB97X-D3/6-31G* level using TeraChem, with $S_1$ treated using TDA [49] and target root 1. The shared geometry pool was partitioned into 800 training, 100 validation, and 100 test configurations. Each potential-energy surface was fitted separately using a force-weighted MACE model. The original test partition was excluded from parameter fitting but was consulted during S1 model refinement and was therefore not reserved solely for final evaluation.

Following $S_0$ optimization of an enol structure, a continuous 1 ns Langevin trajectory was propagated at 300 K using a 0.5 fs timestep. One hundred coordinate–velocity pairs, sampled at 10 ps intervals, initialized switches to the $S_1$ potential. Each switch was followed by 1 ps of unbiased velocity-Verlet dynamics in the NVE ensemble. Proton transfer was defined as the first negative-to-nonnegative crossing of $\Delta_{PT} = d(O - H) - d(N - H)$, and crossing times were determined by linear interpolation between successive 0.5 fs samples. Sampling, model training, trajectory selection and first-passage statistics are detailed in Supplementary Note 9.

**4.9 Predictive accuracy and statistical analysis**

Prediction errors were evaluated against retained QC labels within each specified data partition. For residuals $e_i = \hat{y}_i - y_i$, the metrics were defined as:

$$\text{MAE} = \frac{1}{n}\sum_{i=1}^{n}|e_i|, \qquad \text{RMSE} = \left(\frac{1}{n}\sum_{i=1}^{n} e_i^2\right)^{1/2}, \qquad R^2 = 1 - \frac{\sum_{i=1}^{n} e_i^2}{\sum_{i=1}^{n}(y_i - \bar{y})^2},$$

where $n$ is the number of evaluated predictions and $\bar{y}$ is the reference mean of the evaluated population. Energy errors were reported in eV or explicitly converted to meV, force-component errors in eV $Å^{-1}$, and $R^2$ as a dimensionless quantity. Electronic states, Hamiltonian elements and distance bins were evaluated separately.

Structures, scalar predictions, trajectories and complete workflow executions were treated as distinct reporting units. Trajectory distributions and ensemble-coordinate observables were summarized separately, with censoring and uncertainty procedures specified for each workflow. Saved split assignments and selected checkpoints defined the reported models. Held-out identities alone did not establish statistically independent sampling or an untouched test set. Workflow-specific evaluation provenance and split records are provided in Supplementary Notes 6–10.

**4.10 Reproducibility and computational cost**

Reproducibility records linked configurations, software and hardware snapshots, random seeds, status records and source artifacts. Retries and excluded attempts were distinguished from excluded structures, and artifact completion was distinguished from scientific acceptance.

For the HBQ workflow, computational cost was assessed by multiplying state-specific molecular step counts by mean timings for successful QC gradient evaluations. These serial-equivalent QC estimates were compared with recorded ML dynamics timings. Recorded task timings and serial-equivalent QC estimates were distinguished from calendar wall time and total compute allocation. Complete historical totals for token usage, monetary cost, and human intervention were not established. Seed records, software and hardware information, provenance manifests and timing definitions are provided in Supplementary Note 10.

References

1 Conti, I., Cerullo, G., Nenov, A. & Garavelli, M. Ultrafast spectroscopy of photoactive molecular systems from first principles: Where we stand today and where we are going. *Journal of the American Chemical Society* **142**, 16117 (2020).

2 Curchod, B. F. E. & Martínez, T. J. Ab Initio Nonadiabatic Quantum Molecular Dynamics. *Chemical Reviews* **118**, 3305–3336 (2018). https://doi.org/10.1021/acs.chemrev.7b00423

3 Behler, J. & Parrinello, M. Generalized Neural-Network Representation of High-Dimensional Potential-Energy Surfaces. *Physical Review Letters* **98**, 146401 (2007). https://doi.org/10.1103/PhysRevLett.98.146401

4 Unke, O. T. *et al.* Machine Learning Force Fields. *Chemical Reviews* **121**, 10142–10186 (2021). https://doi.org/10.1021/acs.chemrev.0c01111

5 Chen, C. & Ong, S. P. A universal graph deep learning interatomic potential for the periodic table. *Nature Computational Science* **2**, 718–728 (2022). https://doi.org/10.1038/s43588-022-00349-3

6 Deng, B. *et al.* CHGNet as a pretrained universal neural network potential for charge-informed atomistic modelling. *Nature Machine Intelligence* **5**, 1031–1041 (2023). https://doi.org/10.1038/s42256-023-00716-3

7 Westermayr, J. & Marquetand, P. Machine learning for electronically excited states of molecules. *Chemical Reviews* **121**, 9873–9926 (2021).

8 Dral, P. O. & Barbatti, M. Molecular excited states through a machine learning lens. *Nature Reviews Chemistry* **5**, 388–405 (2021). https://doi.org/10.1038/s41570-021-00278-1

9 Axelrod, S., Shakhnovich, E. & Gómez-Bombarelli, R. Excited state non-adiabatic dynamics of large photoswitchable molecules using a chemically transferable machine learning potential. *Nature Communications* **13**, 3440 (2022). https://doi.org/10.1038/s41467-022-30999-w
10 Müller, C. *et al.* Machine learning for nonadiabatic molecular dynamics: best practices and recent progress. *Chemical Science* **16**, 17542–17567 (2025).
11 Boiko, D. A., MacKnight, R., Kline, B. & Gomes, G. Autonomous chemical research with large language models. *Nature* **624**, 570–578 (2023). https://doi.org/10.1038/s41586-023-06792-0
12 M. Bran, A. *et al.* Augmenting large language models with chemistry tools. *Nature Machine Intelligence* **6**, 525–535 (2024). https://doi.org/10.1038/s42256-024-00832-8
13 Campbell, Q., Cox, S., Medina, J., Watterson, B. & White, A. D. MDCrow: automating molecular dynamics workflows with large language models. *Machine Learning: Science and Technology* **7**, 025037 (2026).
14 Zou, Y. *et al.* El Agente: An autonomous agent for quantum chemistry. *Matter* **8** (2025).
15 Gadde, R. S. K. *et al.* Chatbot-assisted quantum chemistry for explicitly solvated molecules. *Chemical Science* **16**, 3852–3864 (2025).
16 Lahouari, A., Rogal, J. & Tuckerman, M. E. Automated machine learning pipeline: large language models-assisted automated data set generation for training machine-learned interatomic potentials. *Journal of Chemical Theory and Computation* **22**, 305–317 (2026).
17 Kapoor, S. & Narayanan, A. Leakage and the reproducibility crisis in machine-learning-based science. *Patterns* **4** (2023).
18 Liu, K.-L. *et al.* Excited-State Dynamics of [(1, 1′-Biphenyl)-4, 4-diyldi-2, 1-ethenediyl] bis (dimethylsilane). *The Journal of Physical Chemistry A* **113**, 1218–1224 (2009).
19 Fey, M. & Lenssen, J. E. Fast graph representation learning with PyTorch Geometric. *arXiv preprint arXiv:1903.02428* (2019).
20 Schütt, K. T., Sauceda, H. E., Kindermans, P. J., Tkatchenko, A. & Müller, K. R. Schnet–a deep learning architecture for molecules and materials. *The Journal of chemical physics* **148** (2018).
21 Eastman, P. *et al.* OpenMM 7: Rapid development of high performance algorithms for molecular dynamics. *PLoS computational biology* **13**, e1005659 (2017).
22 Seritan, S. *et al.* TeraChem: A graphical processing unit-accelerated electronic structure package for large-scale ab initio molecular dynamics. *Wiley Interdisciplinary Reviews: Computational Molecular Science* **11**, e1494 (2021).
23 Ren, F., Chen, X. & Liu, F. Size-transferable prediction of excited state properties for molecular assemblies with a machine learning exciton model. *The Journal of Physical Chemistry Letters* **16**, 2541 (2025).
24 Li, X., Parrish, R. M., Liu, F., Kokkila Schumacher, S. I. & Martínez, T. J. An ab initio exciton model including charge-transfer excited states. *Journal of chemical theory and computation* **13**, 3493–3504 (2017).
25 Mank, D., Raytchev, M., Amthor, S., Lambert, C. & Fiebig, T. Femtosecond probing of the excited state absorption and structural relaxation in biphenyl derivatives. *Chemical physics letters* **376**, 201–206 (2003).
26 Hruska, E., Gale, A., Huang, X. & Liu, F. AutoSolvate: A toolkit for automating quantum chemistry design and discovery of solvated molecules. *The Journal of Chemical Physics* **156** (2022).
27 Laio, A. & Parrinello, M. Escaping free-energy minima. *Proceedings of the national academy of sciences* **99**, 12562–12566 (2002).
28 Batatia, I., Kovacs, D. P., Simm, G., Ortner, C. & Csányi, G. MACE: Higher order equivariant message passing neural networks for fast and accurate force fields. *Advances in neural information processing systems* **35**, 11423–11436 (2022).
29 Eastman, P. *et al.* OpenMM 8: molecular dynamics simulation with machine learning potentials. *The Journal of Physical Chemistry B* **128**, 109–116 (2024).

30 Case, D. A. *et al.* AmberTools. *Journal of chemical information and modeling* **63**, 6183–6191 (2023).

31 Raucci, U. Capturing excited state proton transfer dynamics with reactive machine learning potentials. *The Journal of Physical Chemistry Letters* **16**, 4900–4906 (2025).

32 Kim, S. *et al.* PubChem 2025 update. *Nucleic acids research* **53**, D1516–D1525 (2025).

33 Grimme, S., Bannwarth, C. & Shushkov, P. A robust and accurate tight-binding quantum chemical method for structures, vibrational frequencies, and noncovalent interactions of large molecular systems parametrized for all spd-block elements (Z= 1–86). *Journal of chemical theory and computation* **13**, 1989–2009 (2017).

34 Tribello, G. A., Bonomi, M., Branduardi, D., Camilloni, C. & Bussi, G. PLUMED 2: New feathers for an old bird. *Computer physics communications* **185**, 604–613 (2014).

35 Invernizzi, M. & Parrinello, M. Rethinking metadynamics: from bias potentials to probability distributions. *The journal of physical chemistry letters* **11**, 2731–2736 (2020).

36 Tully, J. C. Molecular dynamics with electronic transitions. *The Journal of Chemical Physics* **93**, 1061–1071 (1990).

37 Mai, S., Marquetand, P. & González, L. Nonadiabatic dynamics: The SHARC approach. *Wiley Interdisciplinary Reviews: Computational Molecular Science* **8**, e1370 (2018).

38 Westermayr, J., Gastegger, M. & Marquetand, P. Combining SchNet and SHARC: The SchNarc machine learning approach for excited-state dynamics. *The journal of physical chemistry letters* **11**, 3828–3834 (2020).

39 Zhang, Y. *et al.* DP-GEN: A concurrent learning platform for the generation of reliable deep learning based potential energy models. *Computer Physics Communications* **253**, 107206 (2020).

40 Westermayr, J. *et al.* Machine learning enables long time scale molecular photodynamics simulations. *Chemical science* **10**, 8100–8107 (2019).

41 Wang, J., Wolf, R. M., Caldwell, J. W., Kollman, P. A. & Case, D. A. Development and testing of a general amber force field. *Journal of computational chemistry* **25**, 1157–1174 (2004).

42 Becke, A. D. Density-functional thermochemistry. III. The role of exact exchange. *The Journal of chemical physics* **98**, 5648–5652 (1993).

43 Lee, C., Yang, W. & Parr, R. G. Development of the Colle-Salvetti correlation-energy formula into a functional of the electron density. *Physical review B* **37**, 785 (1988).

44 Jakalian, A., Jack, D. B. & Bayly, C. I. Fast, efficient generation of high-quality atomic charges. AM1-BCC model: II. Parameterization and validation. *Journal of computational chemistry* **23**, 1623–1641 (2002).

45 Rohrdanz, M. A., Martins, K. M. & Herbert, J. M. A long-range-corrected density functional that performs well for both ground-state properties and time-dependent density functional theory excitation energies, including charge-transfer excited states. *The Journal of chemical physics* **130** (2009).

46 Hariharan, P. C. & Pople, J. A. The influence of polarization functions on molecular orbital hydrogenation energies. *Theoretica chimica acta* **28**, 213–222 (1973). https://doi.org/10.1007/BF00533485

47 Barducci, A., Bussi, G. & Parrinello, M. Well-tempered metadynamics: a smoothly converging and tunable free-energy method. *Physical review letters* **100**, 020603 (2008).

48 Lin, Y.-S., Li, G.-D., Mao, S.-P. & Chai, J.-D. Long-range corrected hybrid density functionals with improved dispersion corrections. *Journal of Chemical Theory and Computation* **9**, 263–272 (2013).

49 Hirata, S. & Head-Gordon, M. Time-dependent density functional theory within the Tamm–Dancoff approximation. *Chemical Physics Letters* **314**, 291–299 (1999).

# Supplementary Information

PULSES multi-agent AI for autonomous excited-state model building and dynamics

## Table of Contents

## Supplementary Note 1. Benchmarking Framework

The benchmark (Table 1 in the main text) separates three questions: the intrinsic complexity of the requested task (STCI), the scientific capability and causal depth demonstrated in execution (VBD), and accuracy relative to a validated reference execution. The benchmark suite contains 15 cases.

Some scientific problems are intentionally represented by more than one case to isolate workflow scope. For example, benzene p-stacking appears as a submission-only 121-point grid task as well as a complete interaction-surface task. Ethylene appears as an MD sampling/data-handoff task and as an end-to-end MD-QC-ML task. These paired cases share scientific context but have different requested stopping points. Below, we describe each metric in greater detail.

### 1.1 Scientific Task Complexity Index (STCI)

STCI quantifies the scientific demands of each task using the five components defined in Supplementary Table 1. Scores reflect the requirements of the fixed task specification, independently of execution outcome.

**Supplementary Table 1. Components and definitions of the Scientific Task Complexity Index.**

| Component | Definition |
|---|---|
| Workflow Depth (D) | Longest dependency-gated chain of scientific stages. |
| Scientific Domains (S) | Number of distinct scientific domains that must be connected. |
| Scientific Constraints (C) | Independent scientific constraint families. |
| Silent-error Checkpoints (E) | Distinct silent-error checkpoint families. |
| Required Refinement (R) | Ordinal level of refinement required by the task: **0**, single-pass protocol; **1**, fixed comparison or predetermined scan; **2**, adaptive numerical refinement; **3**, model-quality-driven tuning; **4**, diagnostic-driven acquisition of new labels followed by retraining and reevaluation. |

The total score is calculated as:

$$\mathrm{STCI} = 20 \cdot \frac{\min(D, 8)}{8} + 20 \cdot \frac{\min(S, 6)}{6} + 20 \cdot \frac{\min(C, 12)}{12} + 20 \cdot \frac{\min(E, 12)}{12} + 20 \cdot \frac{R}{4}$$

with each component contributes at most 20 points, giving a total score from 0 to 100, rounded to one decimal place.

### 1.2 Verified Breadth-Depth (VBD)

VBD evaluates the scientific capabilities demonstrated within each task's requested scope using archived execution records and validated outputs. It is reported as (B, D), where B measures capability breadth and D measures the depth of connected capabilities. Each capability class in Supplementary Table 2 receives a score of 1 when its evidence requirements are met and 0 otherwise. Each class contributes at most once to breadth, regardless of how many calculations or workflow stages involve that capability. The scoring criteria for B and D are summarized in Supplementary Table 3.

**Supplementary Table 2. Scientific capability classes and evidence requirements for VBD.**

| Class | Evidence Requirement (satisfy=1/unsatisfying=0) |
|---|---|
| CFG(Configurations) | Generation or selection of molecular structures with validated identity and coverage of the declared configuration set. |
| GSQ (Ground-state Quantum Chemistry) | Completed, validated ground-state energies or gradients, accompanied by the relevant method, charge and spin information. |
| ESQ (Excited-state Qunatum Chemistry) | Completed excited-state calculations producing the requested energies, gradients, electronic states or couplings, with appropriate sanity checks. |
| AGG (Scientific Aggregation) | Assembly of individual outputs into a complete, validated scientific product, such as a spectrum, energy surface, Hamiltonian, or curated dataset. |
| MOD (Training a Model) | Production of an identifiable model version with evaluation on a fixed held-out dataset. |
| DEP (Deployment in Optimization or Dynamics) | Perform QM/MM or a ML/MM for validated geometry optimization or dynamics. |
| INF (Physical/Chemical Inference) | Explicit chemical/physical conclusion under a stated criterion, uncertainty or censoring rule. |

**Supplementary Table 3. Calculation of VBD breadth and depth.**

| Label | How it is calculated |
|---|---|
| B (Breadth) | Number of demonstrated capability classes with range from 0 to 7: B = CFG + GSQ + ESQ + AGG + MOD + DEP + INF |
| D (Depth) | Largest number of distinct scientific capabilities linked in the longest validated sequence from the input. Each step must use the output of the preceding step, with evidence confirming that connection. |

## 1.3 Accuracy and reproducibility

Each task was evaluated against a validated reference execution obtained manually without agent assistance. Prompts were refined during development and then held fixed for evaluation. Development runs were excluded from the reported results.

Tasks producing scalar outputs were executed three times, and accuracy was summarized as the mean absolute error (MAE) across the three runs relative to the corresponding reference value. It is worth noting that accuracy denotes agreement with the validated reference execution, rather than the physical accuracy of the underlying electronic-structure method against experiment.

Multistage workflows were assessed using task-specific criteria covering protocol adherence, workflow completion, validity of generated artifacts, and agreement with reference scientific outcomes. Repeated workflows were labelled Pass (3/3) only when all three executions within the specified task scope met these criteria without human intervention. The computationally expensive demonstration workflows were evaluated once and reported as Validated (n = 1). Detailed procedures for reference generation, task-specific acceptance, and run-level benchmark records are provided in the subsections on task-specific acceptance criteria, run-level evaluation records and human intervention below.

## 1.4 Scoring summary

Supplementary Table 4 summarizes the STCI component values, verified capability indicators, and resulting STCI and VBD scores for the 15 benchmark tasks.

**Supplementary Table 4. STCI components and VBD capability profiles across the benchmark suite.**

| Task | STCI components (D,S,C,E,R) | STCI | VBD Components (CFG, GSQ, ESQ, AGG, MOD, DEP, INF) | VBD |
|---|---|---|---|---|
| 1. Benzene (single-point energy) | (2,2,4,4,0) | 25.0 | (1,1,0,0,0,0,0) | (2,2) |
| 2. Benzene (ionization potential) | (3,3,6,5,0) | 35.8 | (1,1,0,1,0,0,1) | (4,4) |
| 3. Benzene (electron affinity) | (4,3,7,6,1) | 46.7 | (1,1,0,1,0,0,1) | (4,4) |
| 4. Benzene ($\omega$ tuning) | (5,3,8,7,2) | 57.5 | (1,1,0,1,0,0,1) | (4,4) |
| 5. Perylene (TDDFT spectroscopy) | (4,3,8,7,1) | 50.0 | (1,0,1,1,0,0,1) | (4,4) |
| 6. Biphenyl (MD) | (7,3,8,8,0) | 54.2 | (1,0,0,1,0,0,1) | (3,3) |
| 7. Benzene π-stacking (submission) | (3,2,6,5,1) | 37.5 | (1,1,0,0,0,0,0) | (2,2) |
| 8. Benzene π-stacking (surface) | (5,4,8,8,1) | 57.5 | (1,1,0,1,0,0,1) | (4,4) |
| 9. Anthracene exciton-model scan | (5,4,10,9,1) | 62.5 | (1,0,1,1,0,0,1) | (4,4) |
| 10. Ethylene MD (sampling) | (6,3,8,8,0) | 51.7 | (1,0,0,1,0,0,0) | (2,2) |
| 11. Ethylene MD-QC-ML | (11,5,15,14,3) | 91.7 | (1,0,1,1,1,0,0) | (4,4) |
| 12. Perylene dimer excitation-energy scan | (5,4,10,9,1) | 62.5 | (1,1,1,1,0,1,0) | (5,4) |
| 13. Perylene Exciton Model Training | (8,6,16,15,4) | 100 | (1,0,1,1,1,0,0) | (4,4) |
| 14. Biphenyl Excited-state Relaxation in Solvents | (14,7,18,17,1) | 85.0 | (1,1,1,1,1,1,1) | (7,6) |
| 15. HBQ ESIPT | (12,7,16,15,3) | 95.0 | (1,1,1,1,1,1,1) | (7,6) |

## 1.5 Task-specific acceptance criteria

Acceptance was evaluated within the stopping scope of each task. A submission-only scan requires validated geometries, calculation inputs, geometry-to-job mapping, and successful submission. A completed scientific surface additionally requires finished calculations, valid parsing, grid completeness, consistent units, and a traceable mapping from geometry to the aggregated observable. Spectra require the specified excited-state quantities, environmental treatment, energy conversion, and broadening. MD-derived datasets require the requested trajectory protocol, conformer count, molecular identity, and coordinate provenance.

For the three anthracene executions, completed grid entries were checked for TeraChem termination, molecular integrity, required method and RESP/ESP input settings, and extraction of an 8 × 8 exciton Hamiltonian and eight adiabatic energies. The central geometry in runs 1 and 3 carried a large-coupling warning and was retained in the accepted surfaces with its flag. Run 2 reported no flagged accepted entries. Model-building and dynamics criteria for the four detailed demonstrations are given in Supplementary Notes 6–9.

## 1.6 Run-level evaluation records

The anthracene evaluation comprises anthracene-1/interactive_1011_213909, anthracene-2/interactive_1012_084551, and anthracene-3/interactive_1012_090051. All three used wPBEh/6-31G* and the requested RESP/ESP settings. Run 1 accepted 289/289 geometries on a 17 × 17 grid with both lateral displacements spanning −4 to 4 Å in 0.5 Å increments at 3.4 Å separation. Run 2 accepted 289/289 geometries on the same lateral grid at 3.5 Å separation. Run 3 accepted 273/273 geometries on a 21 × 13 grid spanning −5 to 5 Å along the long axis and −3 to 3 Å along the short axis, in 0.5 Å increments, at 3.4 Å separation. The three runs therefore document repeated completion of the requested scan workflow with the recorded protocol choices.

The runtime records store one resolved prompt hash for run 1 and a shared hash for runs 2 and 3. Thus prompt identity is preserved at run level alongside the scan definitions. The three anthracene runs identified above are the evaluation executions; the earlier anthracene_pipi_112907 pilot is a separate archive.

The detailed demonstration records are associated with ethylene_md_qc_ml_112915, exmod_revised_131876, the biphenyl_relaxation_107273 model-building run and its nine-solvent production campaign, and hbq_esipt_112917. Retained model and analysis artifacts are identified in Supplementary Notes 6–10. A simulation trajectory, a model-tuning trial, and a recovery attempt are different units from a complete benchmark execution.

## 1.7 Human intervention, provenance, timing, and cost records

Each benchmark run began with a planning stage. The multi-agent system presented its plan, and the user either returned it with clarifications or approved it directly by sending START. Autonomous execution began only after approval. The planner asked questions only where the prompt left a choice open. Examples include whether the 300 s QC time limit applied to each job or to the whole batch (ethylene MD-QC-ML), which torsion-angle convention to report (biphenyl MD), and whether to use full TDDFT or the Tamm–Dancoff approximation (perylene TDDFT). These questions were requests for clarification, not signs that the system had misunderstood the task. Where the prompt already specified these details, no clarification was needed, which is why many runs were approved without any change to the plan. The answers clarified conventions, reporting format or open parameter values, for example asking for a separate gas-phase spectrum alongside the combined plot (perylene TDDFT). They did not add new tasks or change the scientific goal. After approval, nobody gave the system any input, and no session was stopped and resumed by hand. Automatic retries and scheduler continuations count as part of the same run, not as extra repetitions; a run restarted from the beginning was counted as a new run, and the stopped attempt was left out. The prompts for all benchmarks are provided in the SI. Controller elapsed time, summed task durations and model-service usage measure different resources; the production HBQ timing definitions are given in Supplementary Note 10.

## Supplementary Note 2. Exact benchmark prompts

The exact prompts are reproduced below to make the requested workflow scope and stopping condition explicit.

### 2.1 Benzene (single-point)

Run a benzene single-point energy calculation using B3LYP/6-31G*.
The benzene  xyz file is at benzene.xyz.
Submit via SLURM, wait for completion, and report the total energy in Hartree.

### 2.2 Benzene (ionization potential)

Calculate the ionization potential (IP) of benzene using B3LYP/6-31G*.
The benzene xyz file is benzene.xyz.
This requires computing the energy of neutral benzene (charge=0, spinmult=1) and the cation (charge=+1, spinmult=2, unrestricted ub3lyp).
Submit both jobs via SLURM, wait for completion, and report IP = E(cation) - E(neutral) in eV.

### 2.3 Benzene (electron affinity)

Study the effect of diffuse functions on benzene's electron affinity (EA).
The benzene xyz file is in benzene.xyz.
Run calculations with def2-svp and def2-svpd basis sets.
For each basis set, compute neutral benzene (charge=0, spinmult=1, b3lyp) and the anion (charge=-1, spinmult=2, ub3lyp).
EA = E(neutral) - E(anion).
Report all energies and the EA values in eV, comparing the two basis sets.

## 2.4 Benzene (w tuning)

Find the optimal omega value for the wpbeh functional for benzene satisfying the IP-tuning (Koopmans) condition. Use benzene.xyz and the 6-31G* basis set, with wpbeh's default short-range HF exchange partition. Minimize J = |IP + HOMO_energy|, where IP = E(cation) - E(neutral). Use charge=0 spinmult=1 (wpbeh) for the neutral and charge=1 spinmult=2 (uwpbeh) for the cation. Scan omega values from 0.1 to 0.4 in steps of 0.05, then refine around the minimum. Report the optimal omega and the corresponding J value (in eV).

## 2.5 Perylene (TDDFT spectroscopy)

Compute the first 10 excited states of perylene using TDDFT with B3LYP/6-31G*.
The perylene xyz file is at perylene.xyz.
First run the gas-phase calculation with cis=yes and cisnumstates=10, then plot the UV-Vis spectrum with Gaussian broadening sigma=0.2 eV.
Repeat with neq-LR-PCM in acetonitrile and plot both spectra on the same figure as absorption_comparison.png

## 2.6 Biphenyl (MD)

Run a GPU-accelerated molecular dynamics simulation of a single biphenyl molecule solvated by cyclohexane in a 30x30x30 Å box using OpenMM. Then analyze the distribution of the dihedral connecting the two phenyl rings.

Details as follows:

1. There are no structures available yet. Retrieve the relevant structure from PubChem.
2. Build the solvated system using AutoSolvate with default GAFF parameters.
3. Run a minimize-heat-equilibrate-produce workflow with OpenMM, using GPU acceleration if available. Follow this steps:
- minimize
- heat to 300 K in NVT for 100 ps with 1 fs timestep
- run NPT for 100 ps with 1 fs timestep
- run NPT production for 5 ns with 1 fs timestep, saving snapshots every 1000 steps (1 ps)
4. Check the dihedral angle distribution of the biphenyl molecule during the simulation.
5. Report the dihedral angle distribution with proper chart types.

## 2.7 Benzene p-stacking (submission)

Goal: Perform a pi-pi interaction scan on benzene dimer using TeraChem WB97XD3/def2-SVP single point energy, using Slurm, non-dry-run.

Start from an empty workspace and do not assume benzene.xyz is already present. Retrieve benzene coordinates yourself if needed.

Based on benzene, place a second benzene at z=4.0 Angstrom on an x,y grid from -2.0 to 2.0 Angstrom in 0.4 Angstrom steps for 121 total geometries.

Keep all generated files inside the workspace, including any helper CSV, XYZ, or scripts you need.

Then write TeraChem WB97XD3/def2-SVP single-point inputs and SLURM submission assets using the pascal_tc.sh template already in the workspace.

Submit all 121 jobs with sbatch and verify that every task has been submitted.

Stop once all jobs are submitted and report the created files and submitted job identifiers or other submission evidence.

### 2.8 Benzene p-stacking (surface)

Goal: Perform a pi-pi interaction scan on benzene dimer using TeraChem WB97XD3/def2-SVP single point energy, using Slurm, non-dry-run. Compute the interaction energy profile and plot it as a heatmap.

Available concrete input:
- benzene xyz: benzene.xyz

Required outcomes:
- Compute the isolated benzene single-point energy with WB97XD3/def2-SVP.
- Build a benzene dimer scan by placing a second benzene at z=4.0 Angstrom and scanning x,y over [-2.0, -1.6, ..., 1.6, 2.0] with spacing 0.4 Angstrom for 121 total dimers.
- Generate the corresponding XYZ files and a task CSV.
- Use a reusable per-entry QC workflow for the dimer single-point jobs.
- Submit and complete the full 121-entry high-throughput batch via Slurm.
- Compute interaction energies as E_int = E_dimer - 2 * E_monomer.
- Save a summary CSV and a matplotlib heatmap inside the workspace.

Constraints:
- Keep all generated files inside this workspace.
- Do not assume any repo-global imports/parsers are mounted at runtime for the delivered workflow package.
- Use the workflow designer tool and let its evaluator run.
- Do not stop after a pilot subset; all 121 dimer jobs must be submitted and monitored through completion.
- Do not rely on any pre-existing approved workflow asset for this task.

## 2.9 Anthracene exciton-model scan

Compute a 2D pi-stacking PES of anthracene dimers using the TeraChem exciton model using the wPBEh functional and the 6-31g* basis. The anthracene xyz file is at anthracene.xyz. If this file does not exist, stop and report that the input geometry is missing.

The intended PES quantity is exciton-state energy/coupling surfaces.

Add these lines to every TeraChem input:

resp yes

esp_grid_layers 10

esp_grid_dens 2.5

esp_restraint_a 0.0

## 2.10 Ethylene MD (sampling)

Goal: Execute only Step 1 of the ethylene MD sampling flow. Make a directory called "results" to store data in.

Requirements:
- Retrieve ethylene structure and information from PubChem.
- Use AutoSolvate to build a box filled with ethylene and determine the molecule count from density.
- Topology must be created at: results/ethylene_box.prmtop
- Coordinates must be created at: results/ethylene_box.inpcrd
- Heat to 400 K and run NVT. Do NOT use NPT because ethylene is gas.
- Trajectory must be created at: results/openmm_nvt.dcd
- Determine the correct monomer residue name from the topology. If uncertain, inspect residue names and prefer the ethylene residue.
- Sample 1000 monomer conformations.
- Output HDF5 to: results/ethylene_monomer_sample.h5
- Create a QC state CSV from HDF5 at: results/md_state.csv

Stop after Step 1 completes. Do not design QC workflow, do not submit QC jobs, and do not generate any report. Return the exact files created and the residue name used for monomer sampling

## 2.11 Ethylene MD-QC-ML

Goal: Test whether an MD-QC-ML pipeline can be driven end-to-end inside this workspace. Make a directory called "results" to store data in.

a) Stage 1: MD sampling
- Retrieve ethylene structure and metadata from PubChem.
- Use AutoSolvate to build an ethylene box and determine the number of molecules from density.
- Check whether the current node has a GPU. If not, prepare a submission script for GPU-accelerated MD.
- Heat to 400 K and run NVT for 5 ns with 250 step snapshot interval. Do not use NPT because ethylene is gas.
- Determine the correct ethylene monomer residue name from the topology.
- Sample 1000 monomer conformations.
- Produce HDF5 dataset at: results/md_workspace/ethylene_monomer_sample.h5

b) Stage 2: QC calculations
- Compute TDDFT S1-S5 excited-state energies for the ethylene monomer conformations.
- Use B3LYP/6-31G*.
- If the dataset is in HDF5 format, unfold it into XYZ or state inputs as needed before QC.
- The job should end very fast
- Store QC-ready state CSV under: results/qc_workspace/
- You must submit and execute QC for all 1000 conformations, not a pilot subset.
- All jobs should complete within 300 seconds. Verify all calculations succeed before handing off to ML.

c) Stage 3: ML training
- Train a PyTorch Geometric SchNet model on the resulting dataset to fit the ethylene S1 potential-energy surface.
- You may use pretrained initialization or fine-tuning if compatible with the target and data format.
- Use the SchNet guidance from the torch-geometric skill.
- Save ML outputs under: results/ml_workspace/schnet_run/
- Target performance: $R^2$ above 0.99.

d) Stage 4: Fallback planning
- If severe overfitting or underperformance remains after serious hyperparameter tuning, plan dataset expansion.

- The fallback should continue MD sampling with deduplication and continue QC on the new samples.

Keep all work inside this workspace.

### 2.12 Perylene dimer excitation-energy scan

Study excitonic coupling in perylene dimers as a function of stacking distance.
The perylene xyz file is at {{INPUTS_DIR}}/perylene.xyz.
Optimize the monomer, build face-to-face dimers with z-separation from 3.5 to 4.5 Angstrom in steps of 0.1 Angstrom, run both the exciton model and supermolecular TDDFT, and plot S1-S5 energies versus distance in dimer_scan.png.
Use wpbeh functional, include both the first 5 LE states and the CT states in the exciton model, no TT states should be included.

### 2.13 Perylene exciton model training

Mission:
Train ML models to predict all matrix elements of the 4×4 dimer Frenkel exciton Hamiltonian for perylene using only 1 LE state per monomer. All terms must achieve $R^2 > 0.95$ on a held-out test set.

Hamiltonian Structure (4×4)
The dimer Hamiltonian includes:
2 LE energies: E_LE1 (1e(1)g), E_LE2 (2e(1)g, via reversed belonging)
2 CT energies: E_CT1 (1+2-), E_CT2 (1-2+, via reversed belonging)
1 LE-LE coupling: V_LE1-LE2 (1e(1)2e(1))
4 LE-CT couplings:
V_LE1-CT1 (1e(1)1+2-, electron coupling)
V_LE1-CT2 (1e(1)1-2+, hole coupling)
V_LE2-CT1 (2e(1)1+2-, hole coupling reversed)
V_LE2-CT2 (2e(1)1-2+, electron coupling reversed)
CT-CT coupling between reversed CT states is set to zero (as in the paper).
Start execution directly. Do not stop at planning.

Read First:
Read /home/fren5/Exmod_Agent/src/skills/exmod-ml and /home/fren5/Exmod_Agent/src/skills/terachem/references/exciton-model.md for better understanding of the project's background
Optionally read /home/fren5/Exmod_Agent/src/skills/exmod-ml/references/article.md for better understanding of the project's goal and methodology. But do not fully reproduce all methodologies in this article. Focus on the Mission described here.

Interpreted Goal:
Train a machine-learning model for perylene dimers that predicts coupling-related quantities across a broad dimer conformational space.
The key geometric control variable is the nearest heavy-atom distance between the two monomers, spanning approximately 3.5 Angstrom to 10.0 Angstrom.
The final result must not only have strong overall accuracy, but also remain accurate across the entire distance range rather than only in the easy long-range regime.

Fixed Scope:
Do NOT reproduce the full 24000-dimer PrDim workflow exactly.
Do NOT literally reproduce the COM4A / NST5A / SEP5A dataset construction.
In particular, do NOT spend effort reproducing the original SEP5A construction.
If more data is needed later, only do targeted additional sampling / QC in problematic regions.

Do not visit any other folders in the "src" directory except the "skills" folder

Modeling Target:

Preferred target: all non-negligible off-diagonal and diagonal terms (except the CT-CT coupling) that the current ExMod + TeraChem workflow can robustly generate and train.

Minimum acceptable fallback target: at least the LE-LE coupling term must be trained successfully if the full off-diagonal set is not yet robust in the current codebase.

If you must use the fallback target, explain the exact technical reason in the final report and still complete the narrower task successfully.

Execution Plan:

Stage 1: Inspect references and choose the executable target

Responsible Agent: orchestrator, with ml_agent as the primary execution owner for model-scope decisions and qc_agent as the feasibility reference for label availability

Objective: - Read the relevant skill paths. - Validate whether the required scripts exists, especially the exmod_preprocess scripts in the exmod_ml skill.

Required outputs: - A clear chosen prediction target. - A short rationale written into the workspace notes.

Stage 2: Build the initial dimer pool

Responsible Agent: md_agent

Objective: - Build a diverse perylene dimer dataset whose nearest heavy-atom distance covers the full 3.5-10.0 Angstrom range. - Ensure the pool does not collapse onto only close-contact dimers.

Steps: - Use AutoSolvate to construct a 50 A cubic perylene box, simulating condensed phase perylene with its room-temperature density. AutoSolvate supports density-based molecular number calculation and system-construction. - Use OpenMM to run a GPU accelerated MD simulation at 300K for at least 10 ns and 1000 step per snapshot - This trajectory acts as a conformation pool for you to sample. - Required outputs: - A curated MD trajectory for conformation sampling.

Stage 3: Conformation sampling & Define train / validation / uncertainty / test splits

Responsible Agent: md_agent

Objective: - Use the existing target-sampling tool if available. - Construct a held-out uncertainty set spanning the same 3.5-10.0 Angstrom range. - Prefer 0.5 Angstrom distance bins unless a slightly coarser grid is clearly more practical. - Carve out a small official test subset from that held-out uncertainty set.

Requirements: - The uncertainty set must be completely excluded from training. - The official QC-labeled test subset should be about 1000 dimers unless there is a strong technical reason to use slightly fewer. - Ensure train, validation, uncertainty, and test sets all preserve broad distance coverage.

Required outputs: - Split files or manifests that clearly identify the train, validation, uncertainty, and test subsets.

Stage 4: Generate QC labels in bounded batches

Responsible Agent: qc_agent

Objective: - Use the TeraChem Frenkel Exciton Model to generate the labels needed for the ML target. - all QC calculation must be done in wpbeh/6-31G* level. You have to use this DFT functional exactly.

Requirements: - You are strongly recommanded to use high-throughput relevant skills in this task. - Reuse existing ExMod tools and skills instead of inventing a brand-new QC pipeline. - Cache all QC outputs and aggregate them into ML-ready tabular and/or HDF5 datasets. - Keep every label traceable to the source dimer geometry and nearest-heavy-atom-distance metadata. - Do not request an unlimited amount of QC data up front. - Organize QC labeling in bounded batches. - Each QC labeling batch must contain at most 2000 dimers. - For any targeted data-expansion round, prefer adding about 1000-2000 new dimers in that round. - The current QC hardware budget is 20 GPUs total. When launching QC labeling jobs, use a practical concurrent target of 20 GPUs whenever the workflow and scheduler allow it. - Use the 20-GPU concurrency mainly for QC labeling throughput, not for one oversized monolithic job. - You can poll the task status once every hour (3600 seconds).

Required outputs: - Labeled datasets for the initial train/validation split. - Labeled official test subset.

Stage 5: Train the ML model and record tuning history

Responsible Agent: ml_agent

Objective: - Train a model that reproduces the intended coupling-prediction behavior within the current repository.

Requirements: - Reuse the current ExMod ML stack wherever practical. - Read the article and SI to understand the idea, augmented inputs, and matrix-element-specific networks. - Do NOT implement from scratch if the current codebase already has a more robust integrated path. - If uncertainty estimation is not already implemented, use a practical uncertainty proxy such as a deep ensemble across multiple random seeds or another straightforward method. - Assume ML training normally has only one GPU. - Use that GPU as aggressively and efficiently as possible to reduce wall-clock time. - Verify that training is actually using the GPU and avoid accidental CPU fallback. - Do not reserve the 20-GPU QC concurrency budget for ML training. - Before asking for any additional QC data beyond the initial batch, you must complete at least one real training attempt or one meaningful hyperparameter-tuning round. - Before asking for additional QC data, you must leave a persistent tuning record in the workspace containing tried hyperparameters, validation metrics, selected best configuration, and the reason more data is needed. - When more data is added, continue from the best recorded tuning configuration or best checkpoint whenever technically appropriate.

Required outputs: - Training logs. - A saved best checkpoint. - A persistent tuning record in the workspace.

Stage 6: Evaluate, analyze uncertainty, and decide whether targeted resampling is needed

Responsible Agent: ml_agent

Objective: - Evaluate model quality on validation and held-out data. - Identify problematic distance regions.

Required analyses: - Model accuracy versus nearest heavy-atom distance. - Predictive uncertainty versus nearest heavy-atom distance. - Whether high-uncertainty regions align with high-error regions.

Decision rule: - If a distance region fails the required accuracy or shows problematic uncertainty behavior, do targeted additional sampling / QC only in that region, then retrain and reevaluate. - Do not stop at a good average metric that hides a bad distance subregion.

Stage 7: Report final results

Responsible Agent: report_agent

Objective: - Use the report generator to summarize the final model quality and the distance-dependent behavior.

The final report must include at least: - the final model architecture or path actually used - the best checkpoint path and training-output directory - headline validation and test metrics - metric-versus-distance plots - uncertainty-versus-distance plots - if targeted resampling was performed, a before/after comparison - the batch history: initial QC batch size, any additional targeted QC batch sizes, and the tuning stage that justified each extra batch

Success Criteria:

Overall validation-set accuracy must reach $R^2 >= 0.95$.

Overall held-out test-set accuracy must reach $R^2 >= 0.95$.

For every nearest-heavy-atom-distance bin across the validation set and held-out test set, the bin-wise $R^2$ should also be >= 0.95.

If performance remains unsatisfactory after serious model training and tuning, you may request more QC data, but only in targeted problematic regions and only after the tuning record has been written.

Workspace Rules:

Keep all generated files inside the current workspace.

Do not overwrite repository source files unless code changes are truly required for execution.

Prefer organized subfolders such as datasets/, qc_runs/, models/, reports/, and analysis/.

If you encounter a blocking ambiguity, make the most technically reasonable choice, document it, and continue.

Final Deliverables:

A machine-learning-ready labeled dimer dataset with nearest-heavy-atom-distance metadata.

A held-out uncertainty set and a small official test subset.

A trained model and saved best checkpoint.

Evaluation tables and plots.

A final report summarizing accuracy, uncertainty, weak regions, and any targeted resampling performed.

Final Answer Must Report:

where the dataset was written
where the QC labels were written
where the trained model and best checkpoint were written
where the final report was written
whether the per-distance-bin R^2 >= 0.95 requirement was satisfied on both validation and test sets

## 2.14 HBQ ESIPT

Purpose: study the ESIPT dynamics of HBQ using MLIP trained on DFT data.

Background:
The most stable tautomer of HBQ in the ground state is the ENOL (proton on O), but upon photoexcitation to S1 the proton rapidly transfers to N, giving the KETO. we want to study the time for the proton transfer from enol to keto once the molecule is excited to S1.

Step 1: structure prep
- retrieve the 10-hydroxybenzo[h]quinoline (HBQ) structure from PubChem, convert to XYZ, - then identify the index of the important atoms (O, N, transferring H) for later use in defining the CV.

Step 2: conformational sampling using enhanced sampling (GFN-xTB + PLUMED metadynamics)
- use GFN-xTB as a cheap engine to run metadynamics biased along the CV delta_PT = d(O-H) - d(N-H), which captures the proton transfer.
- Here one MUST use a QC sofware that supports xTB + PLUMED, and TeraChem is a good choice.
- One must sample at least 1000 different conformations spanning the enol -> TS -> keto pathway, including the rare TS region, to have good coverage for training. These conformations does not necessarily need to come from one long trajectory.

Step 3: DFT/TDDFT labeling (wb97xd3/6-31G*)
- Use wb97xd3/6-31G* with dispersion enabled.
- For each sampled conformation, run two calculations:
  1. S0 ground-state gradient calculation to get the S0 energy and forces.
  2. S1 excited-state gradient calculation (using TDA) to get the S1 energy and forces.

Step 4: MACE neural-potential training (S0^MLP and S1^MLP)
- Prepare the dataset for MACE
- Train two separate MACE models, one for S0 and one for S1, based on the DFT-labeled energies and forces. Typically Force should be much more important than absolute energy in MLIP simulations, so Force accuracy should reach R2 > 0.99 for a good model.
- If the model failed to reach good accuracy, it's worth trying to evaluate the uncertainty using ensemble methods, then request more data from the region of high uncertainty, and retrain.

Step 5: ESIPT photorelaxation dynamics (using S0^MLP and S1^MLP)
(this task should be deligate to the MD agent who can use OpenMM-ML or ASE-MACE interface, typically MD to run the MLP-driven dynamics, but here is the protocol to follow)
- First, find a ENOL structure (H attached to O) from the dataset as a starting point.
- Optimize the ENOL structure use the S0^MLP to get the reference geometry for the dynamics.
- Run S0^MLP MD of the ENOL structure in gas phase (no solvation needed) at 300K to sample the thermal ensemble. Parameters: timestep 0.5fs, run for 1ns and sample 100 snapshots per 10ps. Each snapshot should include both coordinates and velocities, as they are needed for the Franck-Condon step.
- For each of the 100 snapshots, change the potential from S0 to S1 (this simulates the vertical excitation), and run unbiased NVE dynamics on S1^MLP for 1ps with a timestep of 0.5fs.
- evaluate the proton transfer time by looking at the time evolution of delta_PT and finding when it crosses zero (indicating the proton has transferred from O to N). Average this crossing time over the 100 trajectories to get a mean transfer time, and also look at the distribution of transfer times across trajectories.

Step 6: Report generation
- illustrate the proton transfer dynamics by plotting the average delta_PT(t) curve, and mark the average crossing time.

## 2.15 Biphenyl excited-state relaxation in solvents

Task 1: Train ground- and excited-state MACE models for biphenyl

Execute an end-to-end workflow for biphenyl ground- and excited-state geometry relaxation: generate a solvated box with biphenyl as solute and cyclohexane as solvent using AutoSolvate, equilibrate it with OpenMM, then run PLUMED-OpenMM well-tempered metadynamics with a large bias factor using the biphenyl torsion angle as the collective variable to sample configurations spanning approximately 0 to 180 degrees; extract about 1000 diverse configurations for downstream training, run quantum chemistry labeling on those configurations for both the ground state using `wb97xd3/6-31++G**` and the excited state using TDDFT with the same functional and basis, create 80/10/10 train/validation/test splits, train separate MACE machine-learned PES models for the ground and excited states, and finally use the trained PES models to generate biphenyl geometry-relaxation graphs.

Task 2: Simulate biphenyl excited-state torsional relaxation in explicit solvents

Purpose: Use the trained $S_0$ and $S_1$ MACE models to study biphenyl's structural relaxation following photoexcitation and assess its dependence on solvent environment and viscosity.

Background:

An equilibrated $S_0$ ensemble provides thermally sampled initial structures and velocities. A vertical excitation changes the solute potential from $S_0$ to $S_1$ while preserving the instantaneous nuclear state. Following the subsequent inter-ring motion across many trajectories allows comparison of the relaxation response in different solvents.

Step 1: Prepare the explicit-solvent systems

Reuse the two trained MACE models from Task 1.

Construct periodic systems containing one biphenyl molecule in each of nine solvents: n-hexane, acetonitrile, isooctane, chloroform, methanol, ethanol, 1,4-dioxane, 1-butanol, and n-hexadecane.

Prepare consistent classical solvent parameters and record solvent viscosities at temperatures matching the simulations.

Step 2: Construct and validate the hybrid MACE/MM potential

# Mission: complete the biphenyl S1 solvent-relaxation study and produce a Figure-7-style comparison

Continue the scientific study in the existing workspace:

`/home/ldong/Exmod_Agent/runs/biphenyl_mace_s1_solvent_relaxation_20260821T164306Z_1a56994c`

Use the structures, molecular parameters, system-build products, metadata, provenance records, and other valid scientific assets already present there. Keep all new workflow files, scheduler records, trajectories, analyses, figures, and reports in this workspace. Do not create a replacement campaign or use results, systems, or starting states from other simulation campaigns. The original S0 and S1 MACE models recorded in this workspace are the only inherited external scientific assets.

The goal is to determine how explicit solvent affects the torsional relaxation of plain biphenyl following a vertical S0-to-S1 potential switch, then compare the simulated relaxation timescales with documented experimental solvent viscosities. Study all nine solvents listed in the workspace metadata: n-hexane, acetonitrile, isooctane, chloroform, methanol, ethanol, 1,4-dioxane, 1-butanol, and n-hexadecane.

## Plan and continue from the current state

Begin by auditing the workspace and identifying which preparation, system construction, validation, sampling, and analysis stages are already complete and scientifically usable. Verify important assets from their contents rather than filenames or old status text. Preserve valid completed work and continue from the earliest unfinished or invalid stage.

Based on that audit, write a concise execution plan and proceed with it. Favor the simplest reproducible continuation that uses the existing prepared assets while satisfying the scientific contract in `provenance/contracts/scientific_contract.json`. If additional system assembly or workflow code is needed, derive it from the assets and metadata in this workspace and document the decisions. Do not regenerate structures, molecular parameters, or systems that are already valid.

Use `/home/shared/mamba_envs/exmod_agent` as the scientific runtime. Capture the resolved interpreter, relevant package versions and import paths, OpenMM platforms, CUDA information, and model checksums in the workspace provenance. Validate the actual runtime and model-loading path before production. Resolve any compatibility issue with the smallest scientifically neutral change, preserving the original model files and recording what was done.

## Simulation requirements

Construct and validate an explicit-solvent OpenMM/OpenMM-ML MACE/MM treatment in which the appropriate MACE model supplies biphenyl intramolecular forces and the classical force field describes the solvent and solute-solvent interactions without double counting. Confirm the neutral 22-atom biphenyl ML region, atom ordering, periodic cell, force decomposition, finite energies and forces, S0/S1 switching behavior, and short-time numerical stability for every solvent before production.

Follow the thermodynamic state and sampling target in the existing scientific contract: 298.15 K, 1 bar, and an initial ensemble of 48 independently initiated S1 relaxation trajectories per solvent. Choose and justify stable integration, equilibration, sampling intervals, trajectory duration, output frequency, and restart strategy. The torsion output must resolve the early relaxation dynamics. Use autocorrelation or another defensible diagnostic to assess the effective independence of S0 starting states.

Represent excitation as a vertical switch: retain positions, velocities, and box vectors from each sampled S0 state, replace only the biphenyl S0 potential with the S1 potential, and do not minimize or interpolate on S1 before propagation. Save sufficient trajectory, thermodynamic, torsional, checkpoint, and provenance information to make every result restartable and traceable.

Use Slurm GPU resources efficiently for independent work while respecting current cluster availability. Assess resources immediately before submission, use bounded concurrency, record job IDs and dependencies, monitor submitted work, classify failures, and retry only unfinished transient failures. Never overwrite a valid completed trajectory or replace a failed trajectory with a favorable alternative. If relaxation remains unresolved within the initial observation window, extend sampling according to a documented scientific criterion rather than to improve agreement with viscosity.

## Relaxation analysis

Analyze the central biphenyl torsion as a periodic, symmetry-equivalent coordinate. Use the atom definition already recorded in the workspace, circular/axial statistics where appropriate, and demonstrate branch-cut safety with synthetic tests. Retain signed and folded coordinates for diagnostics.

For each solvent, calculate trajectory-level ensemble relaxation curves and uncertainty bands. Assess whether the dynamics support an identifiable timescale rather than assuming every solvent follows the same decay. Compare reasonable relaxation models, examine residuals and fit-window or plateau sensitivity, and quantify uncertainty with at least 1000 trajectory-level bootstrap replicates. Report censored or non-identifiable behavior honestly.

## Figure-7-style comparison

Create a publication-quality comparison with documented experimental dynamic viscosity near the simulation temperature on the x-axis and the simulated biphenyl S1 torsional relaxation timescale on the y-axis. Include all nine solvents, distinguish hydroxylic protic solvents from the others, show uncertainty and identifiability clearly, and label points legibly. Include a viscosity trend only if the completed results provide adequate statistical support.

This study concerns plain biphenyl torsional relaxation. It does not simulate viscosity, is not an exact reproduction of the experimental substituted-biphenyl Figure 7, and must not be presented as a substituted-biphenyl fluorescence `tau2` measurement.

At minimum, deliver:

- validated system and simulation manifests;
- executable and restartable workflow and Slurm records;
- per-solvent trajectory-status, ensemble, fitting, sensitivity, and bootstrap tables;

- per-solvent relaxation diagnostic figures;
- `figures/figure7_viscosity_vs_relaxation.pdf`;
- `figures/figure7_viscosity_vs_relaxation.png` at 300 dpi or higher;
- `figures/figure7_viscosity_vs_relaxation_data.csv`;
- `figures/figure7_viscosity_vs_relaxation_metadata.json`;
- `reports/final_report.md`; and
- `reports/figure7_completion_summary.json`.

The final report must account for every solvent and every requested trajectory, document runtime and model provenance, explain the chosen methodology and validation evidence, report relaxation estimates with uncertainty and important limitations, and provide exact paths from plotted values back to their source trajectories and metadata. If computation remains active, leave it in a documented restartable state and report the current job IDs and progress rather than claiming completion.

## Supplementary Note 3. PULSES implementation and execution environment

**System prompts.** PULSES uses role-specific system prompts for planning, orchestration, scientific execution, evaluation and reporting. These prompts define each role's responsibilities, tool-use guidance, handoff requirements and expectations for interpreting and reporting evidence. Shared workspace rules and applicable runtime settings are incorporated when the prompts are assembled.

**Model and role assignments.** A Planner supports plan development and revision, while an Orchestrator coordinates five specialist agents for molecular structure preparation, molecular dynamics, quantum chemistry, machine learning and reporting. Additional agents support workflow design and evaluation, while a separate reviewer assesses scientific evidence when needed. Each role is assigned a language model and inference settings, which can be adjusted for individual runs.

**Tool and skill registries.** Tools are explicitly registered for each agent role with defined input schemas. They support molecular preparation, simulation, quantum-chemical calculations, data analysis and reporting, together with general file, Python and shell operations. External software is accessed through libraries, executable programs or service interfaces. Skills complement these tools with domain-specific instructions, reference materials and supporting scripts. Available skills are identified from their metadata and assigned to roles, with relevant skill files copied into the workspace for agents to consult as needed.

**Shared state and artifact schemas.** Tasks enter specialist agents as user messages, and their responses return to the parent as tool messages. Larger artifacts are exchanged through workspace paths. Persistent records separately represent approved plans, runtime configuration, scheduler continuations, scientific audit state and report delivery. QC batch workflows separate task inputs from execution status and use manifests to define input and output requirements. Scientific records link protocols, inputs and results through identifiers and checksums. Saved conversation state supports planning-session restoration and continuation after long-running jobs.

**Software, container and cluster configuration.** The application uses Python, Deep Agents, LangChain and LangGraph, together with scientific libraries and external simulation and quantum-chemistry engines. Development runs used a shared Conda/Mamba environment on the Pascal cluster. The application also supports running in Docker containers. Workspaces use ordinary filesystem access. Computation can run locally or through Slurm, with submission scripts specifying CPU, memory, GPU and wall-time requests. Local worker limits and Slurm array limits control batch concurrency, while job dependencies support deferred continuation.

**Provenance and cost controls.** Run archives retain source snapshots, configuration records, file checksums, scheduler identifiers and execution logs. Plan-dispatch markers, submission identities and locked state transitions reduce duplicate execution during submission and recovery. Shell deadlines, request retries and operation-specific retry guards constrain execution. To reduce unnecessary model usage, scheduler-based continuations resume agent work at relevant workflow stages, allowing long-running calculations to proceed without continuous agent monitoring. Per-call ledgers record model identifiers, roles, timestamps, token usage and errors, supporting retrospective usage accounting. The current implementation does not enforce a task-wide token or monetary budget.

## Supplementary Note 4. Expert-approved planning and the Scientific Execution Contract

**Contract schema.** The Scientific Execution Contract is represented by a Markdown execution plan and a structured approval record. The plan contains three sections: “Interpreted Goal”, “Execution Plan & Agent Allocation”, and “Required Additional Inputs”. Each execution stage specifies the responsible agent, required skills or software, and tasks. The OMWAB components—scientific objective, approved methods, workflow structure, acceptance criteria, and resource and decision boundaries—are expressed within these sections rather than enforced as five mandatory fields in a uniform machine-readable schema. The approval record stores the complete plan, its approval status, content hash, snapshot path and authorization mode.

**Clarification policy.** The Planner develops and revises the plan using the user’s request, subsequent feedback, available agent capabilities, and skill inventory. Its instructions require missing inputs and unresolved scientific choices to remain visible as questions or explicit assumptions for review. User constraints are retained across revisions, and each planning turn presents the complete updated plan. Resolution of scientific questions relies on Planner–user review rather than an independent check by the approval routine.

**Approval policy.** In the current interactive workflow, the user authorizes execution by entering START or its supported Chinese equivalent. The application checks that the plan matches the version presented for review and freezes it without another Planner generation. If the text has changed, the revised draft is presented for review again. The frozen plan is then passed to the Orchestrator. Within its approved scope, agent instructions permit targeted implementation repairs and reuse of unaffected results; changes to scientific scope or requirements require further user approval.

**Plan versioning.** Approved plans are stored as snapshots identified by the SHA-256 hash of their complete text. The application verifies this hash before constructing the execution request, and a startup marker prevents repeated dispatch of the same snapshot through the approval helper. A changed plan has a different content hash; subsequent drafts do not automatically replace the approved version. Stage-level protocol revisions and scheduler continuations are recorded separately and do not constitute renewed approval of the overall contract.

**Final approved contracts.** The workflow-specific Scientific Execution Contracts record the objectives, methods, execution stages, acceptance criteria and decision boundaries agreed through planning and user approval, including accepted assumptions and conditions requiring further approval. These contracts describe the agreed execution requirements; the original task prompts are provided in Supplementary Note 2.

## Supplementary Note 5. Scientific assurance protocols and recovery

**Deterministic validators.** PULSES uses programmatic checks to assess inputs, execution evidence and outputs against applicable protocol requirements. MD validators check data shape and completeness, finite coordinates and physical quantities, structural constraints, simulation duration and specified thermodynamic or sampling criteria, with state-transfer and force-consistency checks applied where required. QC validation includes representative workflow execution, output-availability checks and molecular-integrity checks. ML data utilities check array structure, finite values and split overlap where invoked. These checks produce evidence for scientific assessment; their applicability depends on the domain and stage, and successful execution alone does not establish scientific acceptance.

**Scientific Assurance Agent (SAA) prompts and domain checklists.** Model-based reviewers assess the scientific meaning and adequacy of the available evidence. The MD reviewer's system prompt directs it to evaluate the stated claim against protocol criteria, deterministic findings and supporting summaries, and to identify limitations or required revisions. It receives this evidence in a review packet without file-access or modification tools. The QC evaluator's checklist addresses the appropriateness of calculation settings, units, result extraction, molecular-integrity evidence and workflow robustness; its toolkit permits file and shell operations for inspection and testing. These reviews are separate from the producing agent's assessment, although they need not use a different model family. The implemented review pathways differ by domain, and equivalent independent reviewers are not established for every domain.

**PASS / INCONCLUSIVE / FAIL rules.** These three categories describe acceptance, insufficient evidence and demonstrated failure, respectively, rather than a single shared implementation vocabulary. A PASS applies to the evaluated stage and permitted downstream use. INCONCLUSIVE withholds scientific acceptance when the available evidence or review is insufficient. It does not automatically require repeating the calculation, and downstream use is limited to authorized diagnostic, recovery, or data-generation steps. FAIL denotes a demonstrated violation of applicable requirements, represented by statuses such as MD REJECT or a validator's fail. Native records additionally use REVISE for correctable deficiencies and QC APPROVE for workflow approval. In the MD assurance pathway, blocking deterministic failures cannot be overridden by a favorable model review, while explicitly nonblocking findings may accompany acceptance within a limited claim scope.

**Warning propagation.** Assurance records preserve warnings and limitations together with the assessment decision and permitted use of the results. In the MD pathway, relevant upstream warnings are included in subsequent review evidence. Agent instructions also require unresolved issues to be communicated in downstream handoffs and final reports. Nonblocking warnings allow work to continue while retaining the stated limitations; blocking findings prevent affected results from being treated as accepted inputs. Re-audit or repair does not by itself remove earlier concerns: retained warnings and changes in their status should remain visible in the resulting assessment.

**Revision and assurance-report schemas.** MD result records contain protocol and launch identifiers, phase, deterministic checks, reviewer findings, the combined decision, permitted use and evidence paths with artifact hashes. Protocol revisions identify the preceding protocol, revision type, reason, changed fields and supporting evidence; a re-audit links the new result to the earlier assessment through reaudit_of. QC evaluator reports use a structured Markdown assessment ending in APPROVE, REVISE or REJECT, while archived representative-entry reports use JSON fields for checks, acceptance, evidence and next actions.

These formats preserve the connection between a decision and its evidence without assuming a common report schema for every domain.

**Repair budgets.** Recovery is bounded by the approved scientific scope and operation-specific execution limits. In the current implementation, unchanged MD registration or validation requests are limited to three failed attempts, while an execution retry guard defaults to two failed executions for the same protocol, phase, parameters and parent state. Provider-request retries and QC decision-clarification attempts are separate technical controls, not allowances for scientific revision. The current implementation does not enforce a unified task-wide budget for scientific repairs, resampling, relabelling, retraining or total expenditure. The Orchestrator is instructed to select justified repairs, avoid unchanged failed attempts and report unresolved conditions when further progress requires new information or approval.

**Actual recovery traces.** In ethylene run ethylene_md_qc_ml_112915, the initial QC batch retained 276 valid calculations after a GPU-resource failure. The remaining 724 entries were selectively rerun, yielding complete $S_1$–$S_5$ labels for all 1,000 geometries. The initial SchNet model gave validation and test $R^2$ values of 0.98480 and 0.98490, below the 0.99 target. After 24 additional hyperparameter trials ranked by validation RMSE, the selected model achieved validation and test $R^2$ values of 0.99861 and 0.99630. This trace distinguishes recovery of failed calculations from refinement of a model that did not meet the accuracy target.

## Supplementary Note 6. Ethylene excited-state model construction

The ethylene results correspond to run ethylene_md_qc_ml_112915. Ten ethylene molecules were parameterized with GAFF and AM1-BCC charges, minimized, heated for 50 ps, and sampled using a 5 ns NVT simulation at 400 K with a 1 fs timestep. Coordinates were saved every 250 steps, giving 20,000 frames; 1,000 monomer geometries were selected from source frames separated by 5 ps. TeraChem calculations provided the five lowest singlet vertical excitation energies at the B3LYP/6-31G* level within the Tamm–Dancoff approximation. The model target was $\Delta E_1(R) = E_{s1}(R) - E_{s0}(R)$, the vertical $S_1$ excitation energy at nuclear geometry R, in eV.

A random partition with seed 112915 assigned 800, 100, and 100 geometries to training, validation, and testing. Target normalization used the training-set mean and standard deviation and was inverted before reporting errors in eV. SchNet was trained from scratch. Following an initial baseline, 24 additional hyperparameter trials were evaluated, and trial tune_trial_016 was selected by validation RMSE. Its model used four interaction blocks, 256 hidden channels, 64 filters, 50 Gaussian basis functions, an 8 Å cutoff, a maximum of 32 neighbors, and additive readout. Training used batch size 32, learning rate 0.001753892, weight decay $10^{-6}$, a cosine learning-rate schedule, and seed 112931; the retained checkpoint was epoch 685 of 700.

The retained model had validation and test RMSEs of 0.007197 and 0.010152 eV and $R^2$ values of 0.99861 and 0.99630, respectively. Validation RMSE determined model ranking, while test metrics were recorded during tuning and consulted for workflow acceptance. All partitions originated from the same 400 K sampling campaign, so the reported performance describes prediction within that sampled ensemble.

## Supplementary Note 7. Perylene exciton model construction and evaluation

The perylene model-training results in main-text Section 2.4 and Fig. 3 correspond to campaign exmod_131876. The original development-partition metrics and the subsequent model evaluation are distinguished below.

### 7.1 Model definition and scientific scope

The modelling objective was to predict a perylene dimer's diabatic exciton Hamiltonian from its geometry. Following the exciton-model framework used by Ren et al. (2025), the basis comprised an excitation localized on monomer A, an excitation localized on monomer B, and the two charge-transfer states A → B and B → A. The diagonal entries are the four diabatic state energies. The learned off-diagonal entries comprise one LE–LE coupling and four LE–CT couplings. The real symmetric Hamiltonian therefore contains nine independent learned elements. The coupling between the two oppositely directed CT states was fixed at zero as a modelling approximation; it was not a learned prediction.

The QC Agent generated reference Hamiltonian labels with TeraChem at the ωPBEh/6-31G* level. Preprocessing applied phase correction to establish consistent signs for the coupling targets. Such consistency is required because changes in the arbitrary phases of electronic basis states can change coupling signs without changing the physical system. Each of the nine targets was fitted with a separate neural-network ensemble containing three retained members. Hyperparameter trials preceded expansion, and validation RMSE guided model-configuration and ensemble selection.

### 7.2 Candidate generation and dataset allocation

AutoSolvate and AmberTools prepared a periodic 50 Å cubic box containing 402 perylene molecules, parameterized with GAFF and AM1-BCC charges. OpenMM performed minimization, heating and a 10 ns production simulation at 300 K. The candidate pool actually used for this training campaign contained 13,000 dimers: 10,400 controlled rigid-body configurations and 2,600 dimers extracted from the heating trajectory. Thus, completion of the production simulation does not mean that the training pool was drawn from that production trajectory. A subsequent correction to trajectory selection was not followed by another exciton-model training run; the results reported here describe the original pool. The later 100-dimer production evaluation did not impact model retraining or refinement.

Configurations were grouped by the minimum intermolecular carbon–carbon distance, denoted dmin. Specifically, dmin is the smallest Euclidean separation between a carbon atom on monomer A and a carbon atom on monomer B in the extracted dimer geometry. Thirteen intervals of width 0.5 Å covered 3.5–10 Å. This coordinate measures the closest intermolecular carbon contact and is distinct from the distance between monomer centres of mass. It was used for candidate coverage, training-set expansion and interval-specific accuracy checks. Intermolecular separation provides a useful stratification coordinate because the energies and couplings, and the range of values available for learning and evaluation, vary across intermolecular configurations. It does not describe all orientational or intramolecular degrees of freedom.

The initial allocation comprised 1,600 training dimers, 400 validation dimers, 1,000 test dimers, 1,300 dimers for uncertainty reporting and 8,700 reserve dimers. These counts sum to the 13,000 candidates. The uncertainty set was used to characterize disagreement among the three ensemble members within distance intervals. Reserve selection for the reported expansion used distance-interval checks and a fixed-seed pseudorandom ordering; it was not driven by ranking candidates by ensemble uncertainty.

## 7.3 Accuracy checks and one round of label expansion

The implemented acceptance target was $R^2 \geq 0.95$ overall and within each distance interval for each of the nine Hamiltonian elements. The workflow also assessed the four excitation energies obtained by diagonalizing the assembled Hamiltonian. With 13 targets, 13 intervals and two evaluation partitions, the interval analysis therefore comprised $13 \times 13 \times 2 = 338$ checks. A failed check refers to one target within one distance interval in one partition, rather than to a distinct structural interval. Different targets and partitions can fail in the same interval.

After initial training and hyperparameter trials, 218 of the 338 validation/test interval checks failed, with failures spanning all 13 distance intervals. Model analysis requested 25 reserve candidates per failing interval using a fixed-seed pseudorandom ordering. Dataset preparation supplemented these with another 75 reserve candidates per interval under the supplied expansion rule. Because every interval was affected, this produced 100 additions per interval, or 1,300 additional dimers in total. The QC Agent labelled these existing reserve geometries and the ML Agent retrained the models. No new MD trajectory was generated for this expansion. The operation was an expansion of the labelled training set from a fixed pool, rather than uncertainty-driven acquisition of new configurations.

The final training set contained 2,900 unique dimers. Monomer-swap augmentation produced 5,800 training examples, without doubling the number of independent geometries. The validation and test sets retained their original sizes of 400 and 1,000 dimers. Further model comparisons and validation-based ensemble selection followed the single labelling and retraining round.

The final training set contained 670 dimers at $3.5 \leq d < 5.0$ Å, 1,115 at $5.0 \leq d < 7.5$ Å, and 1,115 at $7.5 \leq d \leq 10.0$ Å.

## 7.4 Final performance and the stopping decision

On the 1,000-dimer test partition, all nine Hamiltonian elements achieved overall $R^2 \geq 0.990$, with target-specific RMSEs spanning 0.484–13.8 meV. The largest target-specific MAE was 7.535 meV. For the four excitation energies obtained by diagonalization, overall test-set $R^2$ values ranged from 0.991 to 0.998 and RMSEs from 2.50 to 13.7 meV. Overall validation and test criteria passed, but 87 of the 338 interval checks remained below $R^2 = 0.95$. Their distribution is summarized in Supplementary Table 5.

**Supplementary Table 5. Residual distance-interval accuracy failures for the perylene exciton model.**

| Target | Validation failures | Test failures |
|---|---|---|
| LE energies and LE–LE coupling | 0 | 0 |
| $CT_1$ energy | 1 | 0 |
| $CT_2$ energy | 1 | 0 |
| $LE_1$–$CT_1$ coupling | 11 | 10 |
| $LE_1$–$CT_2$ coupling | 10 | 10 |
| $LE_2$–$CT_1$ coupling | 10 | 10 |
| $LE_2$–$CT_2$ coupling | 11 | 11 |
| Excitation energies 1 and 2 | 0 | 0 |

| Target | Validation failures | Test failures |
|---|---|---|
| Excitation energy 3 | 1 | 0 |
| Excitation energy 4 | 1 | 0 |
| **Total** | **46** | **41** |

Counts are failed target-by-distance checks among 13 intervals per target and partition. $CT_1$ and $CT_2$ denote A → B and B → A, respectively.

The total is 85 failed Hamiltonian-element checks and two failed excitation-energy checks. All four excitation energies passed in every test interval; the two excitation-energy failures occurred in validation. Failures were not confined to one interval for each target: individual LE–CT couplings failed in 10 or 11 of the 13 intervals in a given partition.

All 87 remaining failed checks had RMSEs below 43 meV, and 78 had reference standard deviations below 1.36 meV. The sensitivity of $R^2$ to the spread of reference values follows from its definition within the evaluated interval:

$$R^2 = 1 - SSE/SST$$

Here, SSE is the sum of squared prediction errors and SST is the sum of squared deviations of reference values from their mean within that same interval. When SST is very small, a small absolute error can produce a low or negative $R^2$. This is a variance effect: the small absolute magnitude of a coupling alone is insufficient to explain low $R^2$. Likewise, low RMSE does not establish that the interval's relative variation has been learned accurately.

On the basis of the absolute-error and reference-variance diagnostics, the Orchestrator declined another expansion and recorded that the strict interval criterion remained unmet. The diagnostics informed the decision to stop further expansion; they did not establish satisfaction of the original interval-specific accuracy requirements. The reported absolute errors do not establish a bound on every configuration's error or certify accurate state mixing and intensities near degeneracies. The observed errors should not be presented as a prospectively specified alternative acceptance rule. The outcome is therefore reported as completion with residual criterion failures, rather than satisfaction of every predefined accuracy requirement.

### 7.5 Evaluation provenance

Training excluded the validation and test dimers, and validation RMSE governed model selection. Nevertheless, selected ensembles were evaluated on both partitions before and after expansion; test evaluation was not conditional on validation passing first. The combined validation/test diagnostics informed training-set expansion and the decision to stop. The test geometries were therefore held out from fitting, but the test results were used during workflow adaptation.

A separate post-development evaluation assessed the ML models on 100 previously unseen dimers from the final 4 ns of one production MD trajectory. Frames were sampled at 40 ps intervals, and one eligible molecular pair was selected uniformly at random from each frame. Eligibility required a minimum intermolecular C–C distance of 3.5–10 Å, with no distance-bin balancing. The resulting set comprised 100 distinct pairs.

The selected models, feature scaling, and 27 retained checkpoints were unchanged from the original development campaign. All 100 configurations were evaluated for each of the nine targets, without offset fitting, observation exclusions, or model selection based on the new results. The trained models achieved RMSEs of 28.2, 30.5, 66.2, and 61.7 meV for $E_{LE1}$, $E_{LE2}$, $E_{CT1}$, and $E_{CT2}$, respectively. The corresponding coupling RMSEs were 1.76, 2.11, 2.53, 2.17 and 1.49 meV for $V_{LE1_LE2}$, $V_{LE1_CT1}$, $V_{LE1_CT2}$, $V_{LE2_CT1}$, and V $V_{LE2_CT2}$. Predictive $R^2$ remained below 0.95 for $E_{LE1}$, $E_{LE2}$, $V_{LE1_CT2}$, and $V_{LE2_CT1}$. Figure 3d summarizes the elementwise RMSEs, with symmetric matrix entries repeating the same metric.

### 7.6 Selected spectra and their reference calculations

Predicted Hamiltonians were diagonalized to obtain excitation energies and the corresponding eigenvectors. For each adiabatic state, its transition dipole was constructed as the linear combination of QC diabatic transition dipoles weighted by the eigenvector coefficients. Oscillator strengths were then obtained from the excitation energies and squared transition-dipole magnitudes. In atomic units, the electric-dipole expression is $f = (2/3)\,\Delta E\,|\mu|^2$. The learned model predicted Hamiltonian elements, rather than transition dipoles. Phase conventions for the dipoles and Hamiltonian must therefore be consistent. Gaussian broadening with a full width at half maximum of 0.060 eV produced the spectra in Fig. 3e.

Figure 3e compares the QC reference and ML predictions for the matched production ensembles. The monomer QC reference comprises the 200 constituent monomer geometries of the 100 selected dimers. Its ML monomer proxy combines the predicted $E_{LE1}$ and $E_{LE2}$ energies with QC monomer oscillator strengths. The proxy energy MAE is 23.2 meV relative to the matched monomer QC excitation energies. The dimer comparison uses all four exciton-model states for each of the same 100 geometries; its energy MAE and RMSE are 36.7 and 49.1 meV relative to QC exciton-model excitation energies.

### 7.7 Reference for the exciton-model framework

Ren, F., Chen, X. & Liu, F. Size-Transferable Prediction of Excited State Properties for Molecular Assemblies with a Machine Learning Exciton Model. J. Phys. Chem. Lett. 16, 2541–2552 (2025). https://doi.org/10.1021/acs.jpclett.4c03548

## Supplementary Note 8. Biphenyl excited-state relaxation in solvents

**Training configurations and QC labels.** The biphenyl models originated from run biphenyl_relaxation_107273. Well-tempered metadynamics in explicit cyclohexane used OpenMM and PLUMED at 300 K and 1 bar with a 1 fs timestep. The biased coordinate was the central four-atom dihedral defined by atoms 1–6–7–8 in the archived topology. Hills of initial height 1.2 kJ $mol^{-1}$ and width 0.24271 rad were deposited every 500 steps with a bias factor of 50. A 200 ps trajectory provided 2,000 frames at 0.1 ps intervals. Selection across eighteen 10° bins of the absolute dihedral over 0–180° retained 1,000 geometries, with 55–56 per bin, using farthest-point selection of Kabsch-aligned heavy-atom coordinates and a 0.02 Å duplicate-filtering threshold.

Paired $S_0/S_1$ energies and gradients were calculated for isolated, neutral singlet biphenyl at the ωB97X-D3/6-31++G** level. The $S_1$ TeraChem inputs used the Tamm–Dancoff approximation, target root 1, and ten requested singlet states. The $S_1$ learning target was the total excited-state energy. Energies were converted from Hartree to eV, and forces were obtained as negative gradients in eV $Å^{-1}$. Both states used the same final torsion-stratified partition of 800 training, 100 validation, and 100 test geometries.

Separate ScaleShiftMACE models used 64 channels, two interaction blocks, correlation order 2, maximum angular momentum L = 1, eight radial basis functions, five cutoff basis functions, and a 5 Å cutoff. The recorded production settings used batch size 64, validation batch size 128, initial learning rate 0.005, energy and force loss weights of 50 and 100, a maximum of 200 epochs, patience 40, evaluation every two epochs, float32 precision, and training seed 20260526. The test force-component RMSEs were 0.0243 and 0.1295 eV $Å^{-1}$ for $S_0$ and $S_1$, respectively.

**Explicit-solvent dynamics.** The nine-solvent production campaign used the $S_0/S_1$ models from biphenyl_relaxation_107273 in n-hexane, acetonitrile, isooctane, chloroform, methanol, ethanol, 1,4-dioxane, 1-butanol, and n-hexadecane. OpenMM-ML replaced classical intramolecular interactions within the 22-atom solute with the corresponding MACE potential, retaining classical solvent and solute–solvent interactions and the same solute charges across the vertical switch. Ground-state preparation comprised 20 ps of NPT equilibration at 298.15 K and 1 bar followed by 192 ps of sampling. Coordinate–velocity–box states saved every 4 ps supplied 48 starting conditions per solvent.

Each vertical $S_0$-to-$S_1$ switch preserved coordinates, velocities, and box vectors without minimization or velocity reassignment. $S_1$ trajectories were propagated for 100 ps with a Langevin-middle integrator, a 0.5 fs timestep, and friction coefficient 1 $ps^{-1}$. Pressure coupling was retained with a Monte Carlo barostat at 1 bar and 298.15 K, with trial moves every 25 steps. The mixed system used particle mesh Ewald electrostatics and a 1.0 nm nonbonded cutoff. Inter-ring angles were recorded every 10 fs, giving 432 trajectories across the nine solvents.

**Relaxation observable.** Dynamics were analyzed using $\theta(t) = \arccos(|nA(t)\cdot nB(t)|)$, where nA and nB are unit normals to least-squares planes fitted to the six carbons of each ring. This coordinate ranges from 0° for coplanar rings to 90° for orthogonal rings. For each trajectory, θstart was the first saved angle at 0.01 ps and θlate was the median angle over the final 10 ps. The threshold was θthresh = θlate + (θstart − θlate)/e. The individual threshold time was the first sampled crossing of this threshold, without interpolation between the 10 fs samples. Solvent-level values were arithmetic means over the 48 trajectories. This threshold statistic does not require fitting an exponential decay.

Torsion-resolved energy errors in Figure 4c were evaluated using the final MACE test partition of 100 geometries. Both electronic states used the same geometry identifiers, with training and validation geometries excluded. Configurations were assigned to eighteen 10° bins of the absolute signed dihedral |φ| over 0–180°, without symmetry folding, giving 5–6 test geometries per bin. For each bootstrap resample, the geometry identifiers within a bin were sampled with replacement, using identical indices for $S_0$ and $S_1$, and RMSE was calculated from the total molecular energy residuals. The plotted center is the mean bootstrap RMSE, and the band spans one sample standard deviation of the bootstrap RMSE estimates. Energy errors were calculated without fitted offsets or per-atom normalization.

The bootstrap bands describe resampling variability of the within-bin RMSE; they are not confidence intervals or variability among independently trained models. Resampling treats geometries as exchangeable and does not account for possible correlations inherited from trajectory sampling. Other internal coordinates vary between configurations, so the panel describes torsion-resolved prediction errors rather than a torsional potential-energy scan.

For Figure 4e, precomputed solvent-level mean threshold times were matched to tabulated dynamic viscosities by solvent identifier. A descriptive relationship, $\bar{\tau} = a\eta + b$, was fitted by unweighted ordinary least squares with a free intercept using the six aprotic solvents. Here $\eta$ denotes dynamic viscosity and $\bar{\tau}$ the arithmetic mean individual-trajectory threshold time for a solvent. The three protic solvents were displayed without contributing to the fit. $R^2$ was calculated from residual and total sums of squares over the six aprotic-solvent means. Uncertainty bars were omitted, and existing uncertainty estimates were not used as regression weights.

## Supplementary Note 9. Excited-state proton transfer in HBQ

The HBQ results correspond to campaign hbq_esipt_112917. Gas-phase TeraChem GFN-xTB dynamics with PLUMED well-tempered metadynamics generated 2,750 frames from twelve retained trajectory segments. Geometry filtering retained 2,717 frames, and deduplication within proton-coordinate bins produced 1,496 candidates. Coordinate-stratified selection retained 1,000 geometries: 350 enol, 300 intermediate, and 350 keto structures. The proton-transfer coordinate was $\delta PT = r(O\text{–}H) - r(N\text{–}H)$, with zero-based atom indices O = 0, N = 1, and transferring H = 17; distances are in Å. Negative values place the hydrogen closer to O and positive values closer to N. The intermediate selection region was $-0.25$ Å $\leq \delta PT \leq 0.25$ Å.

Each geometry received paired $S_0/S_1$ energies and gradients at the ωB97X-D3/6-31G* level. $S_1$ calculations used the Tamm–Dancoff approximation, target root 1, and five requested singlet states. Energies were converted to eV, and forces were obtained from gradients using $F = -\nabla E \times 27.211386245988/0.529177210903$ for gradients in Hartree $bohr^{-1}$. A shared 800/100/100 training/validation/test partition used twenty equal-count proton-coordinate strata together with region labels and seed 20261129.

Figure 5b shows the complete QC-labelled dataset of 1,000 geometries, counted once per geometry: 350 enol, 300 intermediate, and 350 keto structures. The proton-coordinate histogram uses 0.10 Å bins. Counts are unsmoothed and include the tails. These selected counts describe coverage of the labelled dataset rather than equilibrium populations.

Separate ScaleShiftMACE models used 64 channels, two interactions, correlation order 2, maximum angular momentum L = 1, eight radial basis functions, five cutoff basis functions, and a 5 Å cutoff. The selected $S_0$ run, prod_s0, used batch size 64, initial learning rate 0.005, energy/force loss weights 100/500, and seed 20261129; its retained checkpoint was epoch 215. The selected $S_1$ run, prod_s1_b32_f1000_lr003, used batch size 32, initial learning rate 0.003, energy/force weights 100/1000, and seed 20261131; its retained checkpoint was epoch 295. Both final models exceeded force-component $R^2 = 0.99$ on the validation and test partitions. $S_1$ refinement consulted test performance, so the test results describe an adaptively developed model. For Figure 5c, test-set energy errors were calculated after subtracting state-specific mean signed residuals estimated from the validation partition. Force errors were evaluated without correction. The corrected energy MAEs were 0.0246 eV for $S_0$ and 0.0241 eV for $S_1$, while the corresponding raw energy MAEs are 0.080 and 0.123 eV.

$S_0$ dynamics began from an enol geometry optimized with the selected $S_0$ MACE model. A single continuous 1 ns Langevin trajectory at 300 K used a 0.5 fs timestep and friction coefficient 1 $ps^{-1}$. One hundred coordinate–velocity states, saved every 10 ps, initialized vertical switches to the $S_1$ model. Each state was propagated for 1 ps using unbiased velocity-Verlet NVE dynamics with a 0.5 fs timestep. The final production ensemble used this continuous $S_0$ trajectory; the earlier replica-aggregate archive was not included.

All 100 $S_1$ trajectories completed and crossed $\delta PT = 0$ within 1 ps. Individual first-passage times were defined by the first negative-to-nonnegative crossing, linearly interpolated between adjacent 0.5 fs samples. For descriptive analysis, the fast-transfer subset contained the 75 trajectories with individual first-passage times below 100 fs; its conditional mean was 58.25 fs. The remaining 25 trajectories were retained in the full-ensemble statistics, giving a mean of 125.55 fs and median of 70.69 fs over all 100 trajectories. The cumulative first-passage distribution is $F(t) = N^{-1}\Sigma_i I(t_i \leq t)$, with $N = 100$ and I equal to one when the condition is satisfied and zero otherwise. The first zero crossing of the ensemble-mean coordinate occurred at 67.47 fs and is a distinct statistic from the mean of individual crossing times.

## Supplementary Note 10. Cross-workflow reproducibility and resources

**Metric aggregation.** For reference values $y_i$, predictions $\hat{y}_i$, and residuals $e_i = \hat{y}_i - y_i$, MAE is the arithmetic mean of $|e_i|$, RMSE is the square root of the arithmetic mean of $e_i^2$, and $R^2 = 1 - \Sigma_i e_i^2/\Sigma_i(y_i - \bar{y})^2$, where $\bar{y}$ is the reference mean within the evaluated partition. Metrics are calculated separately for each target and evaluation set. Energy errors use eV per molecule unless explicitly converted to meV. Force metrics pool Cartesian components across atoms and geometries within each state and partition. The 100-geometry HBQ test partition therefore contains 7,200 force components per electronic state.

For perylene, predictions are averaged across ensemble members before scoring, and metrics are reported overall and within 0.5 Å distance bins. Overall scores are calculated from individual predictions, not by averaging bin-wise $R^2$. Absolute errors accompany $R^2$ because small reference variance can produce large negative values despite small errors. Neural ensemble dispersion uses sample standard deviation and does not represent a confidence interval for predictive performance. Figure 3d instead reports per-target RMSE over all 100 production dimers without distance-bin averaging or offset correction. Symmetric matrix cells repeat a single target metric and are not independent measurements. Biphenyl Figure 4c uses the mean and sample standard deviation of within-bin bootstrap energy RMSE estimates, with paired geometry resampling across $S_0$ and $S_1$. The solvent-viscosity panel displays solvent-level means of individual-trajectory threshold times without uncertainty bars.

**Seed and split records.** Split manifests retain dataset membership and, where available, source-frame and molecular-pair identifiers. In the perylene ExMod campaign exmod_131876, candidate generation used seed 131876 and distance-stratified split allocation used seed 131877. The initial training, validation, and test sets contained 1,600, 400, and 1,000 dimers, respectively; one reserve-pool expansion increased training to 2,900 while preserving validation and test membership. Initial neural ensembles used seeds 200-202. Final three-member ensembles used seeds 1200-1202 for seven Hamiltonian elements; validation-based selection retained seeds 1203, 1208, and 1207 for $E_{LE1}$ and 1208, 1206, and 1207 for $E_{LE2}$. This campaign did not use the descriptor-bootstrap ensembles from the older perylene run. Validation and test errors informed expansion and stopping decisions, so the final test evaluation describes an adaptively developed campaign rather than an untouched prospective assessment. The additional 100-dimer production cohort was evaluated using the same selected ensembles, with no further fitting or selection. Its results are reported separately from the original 1,000-dimer development-consulted test partition.

For biphenyl, the original $S_0/S_1$ models from biphenyl_relaxation_107273 used training seed 20260526. The dataset-preparation script specifies the same seed for the final torsion-stratified 800/100/100 allocation; retained geometry identifiers define the partition used for evaluation. These models were deployed in the nine-solvent campaign. Its $S_0$ driver initialized velocities with a deterministic solvent-dependent seed, calculated as 100000 plus the sum of each solvent-identifier character's Unicode value multiplied by its one-based position. $S_1$ trajectories inherited saved $S_0$ positions, velocities, and box vectors. The inspected propagation scripts did not explicitly set Langevin integrator seeds. The 48 initial states per solvent were sampled at 4 ps intervals from a common $S_0$ trajectory. Recorded seeds and disjoint dataset identities do not establish statistical independence or guarantee bitwise reproducibility across software and hardware environments.

For HBQ campaign hbq_esipt_112917, the shared S0/S1 geometry split contained 800 training, 100 validation, and 100 test structures, stratified by the proton-transfer coordinate and region using seed 20261129; the selected S0 and S1 models used training seeds 20261129 and 20261131, respectively. S1 model refinement also consulted test performance. HBQ S0 velocity initialization used NumPy seed 2026112906, whereas the inspected driver did not explicitly seed the PyTorch thermostat noise.

**Software and hardware versions.** HBQ training records identify MACE 0.3.16. The final HBQ $S_0$ and $S_1$ dynamics status files identify PyTorch 2.6.0+cu124, CUDA 12.4, and an NVIDIA RTX A5000 GPU. Successful HBQ QC-labeling timings include calculations on RTX A5000 and RTX A4000 GPUs. The biphenyl production training records identify MACE 0.3.16 and PyTorch 2.6.0+cu124. Software versions and hardware are associated with individual stages; later development-environment snapshots do not replace the production-stage records.

**Run configuration records.** The three anthracene evaluation archives retain benchmark_runtime_profile.json and benchmark_events.jsonl. All three runtime profiles record the full configuration, with planning, multiple agents, skills, excited-state guidance, and the independent evaluator enabled, and trial_seed = 0. The profiles retain the resolved prompt hashes, while execution plans and scientific outputs record the run-specific scan choices. These profile fields identify the enabled workflow components and should be distinguished from the role-specific language-model settings described in Supplementary Note 3.

**Computational timing.** The HBQ controller record reports an elapsed span of 17 h 16 min 49 s. Separately, the final dynamics timers record 33,666.89 s for $S_0$ and 107.88 s for the batched $S_1$ ensemble, totaling 9.3819 h. Summing successful paired-label driver times, all six recorded training-log spans, and final production dynamics gives 22.750 h. This partial sum includes overlapping tasks and differs from controller wall time. It excludes unaccounted sampling, failed attempts, queue waiting, analysis, and model-service costs.

For the direct-QC cost estimate, the number of molecular integration steps was multiplied by the mean processing time of a successful QC gradient evaluation for each state. The $S_0$ workload contains 2,000,000 steps and the $S_1$ workload 200,000 steps. Mean QC times of 6.36817 s for $S_0$ and 35.07716 s for $S_1$ give a combined serial-equivalent estimate of 5,486.60 h. This extrapolation assumes one gradient evaluation per molecular step and is separate from measured ML elapsed time and human labor. Recorded timing categories are not interchangeable, and the partial workflow sum does not establish a measured end-to-end speedup.

**Provenance manifests.** The ethylene archive links splits.json, preprocessing_metadata.json, tuning_selection.json, final_metrics.json, and final_schnet_best.pt. The HBQ model_selection_metadata.json identifies the retained $S_0$ and $S_1$ runs, checkpoint epochs, final compiled-model paths, evaluation metrics, and model hashes; the dynamics status files identify the models and initial-state files used in production. Biphenyl mixed-system scripts and trajectory result files retain state-specific model hashes and starting-state references. Anthracene aggregation records connect grid rows to completed QC outputs and the plotted energy and coupling surfaces. These records distinguish selected results from intermediate candidates, retries, and earlier campaigns.

## Supplementary Note 11. Data and code availability and declarations

### 11.1 Data availability

The data supporting this study are planned to be made publicly available upon journal publication. Repository details and accession identifiers will be added when the data are released.

### 11.2 Code availability

The code supporting this study is planned to be made publicly available upon journal publication. The repository location and release identifier will be added when the code is released.

### 11.3 Correspondence

Correspondence should be addressed to Fang Liu (fang.liu@emory.edu).

### 11.4 AI assistance in manuscript preparation

ChatGPT and Codex were used to assist with drafting, revising and editing the manuscript text.